\RequirePackage{ifpdf}
\documentclass[hyper,12pt,letterpaper]{JHEP3}
\pdfoutput=1
\usepackage{graphicx}
\usepackage{amsmath,amssymb,multirow,array}

\numberwithin{equation}{section}

\numberwithin{equation}{section}
\newcommand{\ben}{\begin{eqnarray}\displaystyle}
\newcommand{\een}{\end{eqnarray}}
\newcommand{\be}{\begin{equation}}

\newcommand{\ee}{\end{equation}}
\newcommand{\lb}{\left (}
\newcommand{\rb}{\right )}
\newcommand{\ltb}{\left [}
\newcommand{\rtb}{\right ]}
\newcommand{\ra}{\rightarrow}

\newcommand{\pa}{\partial}

\newcommand{\bc}{\begin{center}}
\newcommand{\ec}{\end{center}}

\newcommand{\M}{ {\cal M}}

\newcommand{\Rmnum}[1]{\expandafter\@slowromancap\romannumeral #1@}

\newcommand{\df}{{\mathrm{d}}}
\newcommand{\e}{{\mathrm{e}}}

\def\cg2{\cos (\pi V)}
\def\sg2{\sin (\pi V)}
\def\cb2{\cos (\delta/2)}
\def\sb2{\sin (\delta/2)}

\def\sg{r_0^2 {\rm sinh}^2\gamma }

\def\cg{r_0^2 {\rm cosh}^2\gamma }

\def\[{\left [}
\def\]{\right ]}
\def\({\left (}
\def\){\right )}

\usepackage{float}
\usepackage[font=small,labelfont=bf]{caption}
\usepackage{mathrsfs}
\usepackage[parfill]{parskip}
\usepackage{subcaption}

\title{Dyonic Black Hole and Holography} \author{Suvankar
  Dutta\footnote{suvankar@iiserb.ac.in}, Akash
  Jain\footnote{akash@iiserb.ac.in, ajainphysics@gmail.com} and Rahul
  Soni\footnote{rahulsoni@iiserb.ac.in, rahulsoni.phy@gmail.com}\\
  Department of Physics \\
  Indian Institute of Science Education and Research Bhopal \\
  Bhopal 462 023\\
  India}

\abstract{We study thermodynamic properties of dyonic black hole and
  its dual field theory. We observe that the phase diagram of a dyonic
  black hole in constant electric potential and magnetic charge
  ensemble is similar to that of a Van der Waals fluid with chemical
  potential. Phase transitions and other critical phenomena have been
  studied in presence of magnetic charge and chemical potential. We
  also analyse magnetic properties of dual conformal field theory and
  observe a $ferromagnetic$ like behavior of boundary theory when the
  external magnetic field vanishes. Finally, we compute susceptibility
  of different phases of boundary $CFT$ and find that, depending on
  the strength of the external magnetic field and temperature, 
  these phases are either paramagnetic or diamagnetic.}  

\begin{document}

\maketitle

\newpage

\section{Introduction and summary}

Recently there has been much interest in holographic study of
condensed matter systems. Different properties of $(d+1)$ dimensional
condensed matter systems (field theories) are being studied from the
perspective of string theory. In this paper we attempt to understand
magnetic properties of a $(2+1)$ dimensional system via the $AdS/CFT$
correspondence (proposed by Maldacena in 1997
\cite{Maldacena:1997re,Witten:1998qj}). For that, we consider a
$(3+1)$ dimensional $dyonic$ black hole spacetime as our bulk system.

Thermodynamic properties of electrically charged black holes in
arbitrary dimensions are well studied in the literature. However, in
four spacetime dimensions, because of electromagnetic duality, it is
possible to construct a black hole which carries both electric and
magnetic charges. Such black hole solution is called dyonic black
hole. In this work we consider a dyonic black hole in asymptotically
$AdS$ spacetime, as we aim to understand thermodynamic properties of
its dual boundary $CFT$.

A dyonic black hole sources two different fields: graviton (metric)
and a $U(1)$ gauge field. The metric satisfies asymptotic $AdS$
boundary conditions and for gauge field there are two possible
fall-off conditions. A normalizable mode that corresponds to a VEV of
the dual operator (an $R$-current), and a non-normalizable mode
corresponding to the application of an external gauge field, which
deforms the boundary theory. The deformed $(2+1)$ dimensional field
theories have drawn enough attention recently in the context of a
holographic understanding of condensed matter phenomena like
superconductivity/superfluidity \cite{hhh}, the Hall effect
\cite{hartnoll-kovtun}, magnetohydrodynamics \cite{maghydro}, the
Nernst effect \cite{nernst} and more.

On the other hand in the bulk theory, presence of a magnetic monopole
enriches the phase diagram of black holes in $AdS$ spacetime.  In
\cite{myers}, the authors studied electrically charged $AdS$ black
holes in diverse dimensions. They found a seemingly surprising
similarity between phase diagrams of black hole and Van der Waals
fluid. Considering the black hole in a fixed electric charge ensemble
it was observed that identifying $\beta, q$ and $r_+$ (inverse
temperature, electric charge and horizon radius of black hole
respectively) with the pressure, temperature and volume of liquid-gas
respectively, black hole phase diagram is similar to that of a
non-ideal fluid described by the Van der Waals equation. Below a
critical value of electric charge, there exists three different
phases/branches of black hole. These three different branches have
different horizon radii. The smallest one is identified with the
liquid state and the largest one is identified with the gaseous state
of Van der Waals fluid. Whereas, the medium size unstable black hole
is identified with the un-physical state of Van der Waals fluid. In
stead of fixed charge ensemble, if we consider the black hole in fixed
electric potential ensemble, the similarity between the phase diagrams
goes away.

The same system has also been studied by \cite{pvcritic}\footnote{See
  also \cite{Altamirano:2013uqa}.}. In this interesting paper the
authors identified the negative cosmological constant with
thermodynamic pressure of the black hole and volume covered by the
event horizon to be the thermodynamic volume conjugate to pressure
(following \cite{kastor}). Under this consideration, the phase diagram
of black hole in canonical ensemble exactly matches with the phase
diagram of the Van der Waals fluid (no need to make any ad-hoc
identification between the parameters like \cite{myers}). But, in this
picture the black hole phase diagram has an extra parameter: electric
charge. Therefore, phase transitions and other critical phenomena
depend on this extra parameter. However, in grand canonical ensemble
the similarity between the phase diagrams is lost again. Higher
derivative generalization of black hole phase diagram has been studied
in \cite{Chen:2013ce}.

We consider a dyonic black hole in $(3+1)$ dimensions and find that
black hole phase diagram is more colorful. We observe that putting the
black hole in constant electric and magnetic charge ensemble one gets
the same result of \cite{myers,pvcritic} only $q_E^2 \ra q_E^2
+q_M^2$. However, if we consider the black hole in constant electric
potential ensemble (keeping magnetic charge arbitrary) then unlike
\cite{myers, pvcritic}, we get a phase diagram which is same as
liquid-gas phase diagram with a chemical potential. Therefore, in this
case phase transitions and different critical behaviour depend on
temperature, volume, pressure, magnetic charge and the electric
potential. In other words, the black hole phase diagram is more rich
and our aim is to explore different parts of this phase diagram which
have not been studied before. We summarize our results below in
section \ref{catastrophic}.

In the context of the $AdS/CFT$ correspondence, the dyonic black hole
is dual description of a $(2+1)$ dimensional $CFT$ with a conserved
$U(1)$ charge ($\sim q_E$) and in a constant magnetic field ($\sim
q_M$). Field theory in presence of external electromagnetic field
displays many interesting behaviours. The second part of this paper is
devoted to study magnetic properties of different phases of the
boundary $CFT$.  We find a \textit{`ferromagnetic'} like behaviour
of boundary theory. We observe that in the limit $B\ra 0$ the
magnetization of the system vanishes beyond a critical temperature,
whereas below that temperature it attains a constant value. However,
unlike ferromagnetism there is a sharp jump in the magnetization at
the critical value. 

\subsection{Summary of main results}

Here we summarize the main results of this paper.

\subsubsection{Black hole phase diagram}\label{catastrophic}

We study thermodynamic properties of dyonic black hole in constant
electric potential and magnetic charge ensemble.  Identifying
$-\Lambda$ as the thermodynamic pressure and its dual as thermodynamic
volume, we see that the system shows almost similar phase diagram as
in \cite{pvcritic}, but its parameter space is more rich.

The phase diagram of dyonic black hole in constant electric potential
has been depicted in figure (\ref{fig:PV}) and figure
(\ref{fig:tpv}). Here we see that for a fixed $\Phi_E<1$ and
$q_M<q_{M(c)}$ there exists a critical temperature $T_c$ below which
the black hole has three different phases inside a particular window
along $P$ axis. We call them {\it small black hole} (SBH), {\it medium
  black hole} and {\it large black hole} (LBH) (according to size
$V$). This phase diagram shows a qualitative similarity with
liquid-gas phase diagram governed by the Van der Waals
equation. However, the black hole equation of state (see equation
(\ref{E:EOS2})) is different than the Van der Waals equation. We shall
see in section \ref{phasestructure} that unlike Van der Waals fluids,
black hole phase diagram is controlled by two extra parameters
$\Phi_E$ and $q_M$ (apart from usual $P$, $T$ and $V$). Hence,
critical points and phase transitions in this phase diagram depend on
the values of these two extra parameters as well.

We calculate the free energy of this system and observe that inside some
region in parameter space ($q_M-\Phi_E$ plane), there exists a phase
transition between large black hole to small black hole as we decrease
temperature. High temperature phase is governed by large black hole
whereas low temperature phase is dominated by small black hole. If we
are outside that region then the only thermodynamically stable phase
is the large black hole phase.

In $q_M \ra 0$ limit, the large black hole to small black hole phase
transition reduces to the Hawking-Page phase transition. In this case
the small black hole evaporates and the low temperature phase is
dominated by a global $AdS$.

\subsubsection{Holography with dyonic black hole: Magnetic properties
  of boundary theory}\label{magprop}
\begin{figure}[h]
	\centering
	\includegraphics[width=0.5\textwidth]{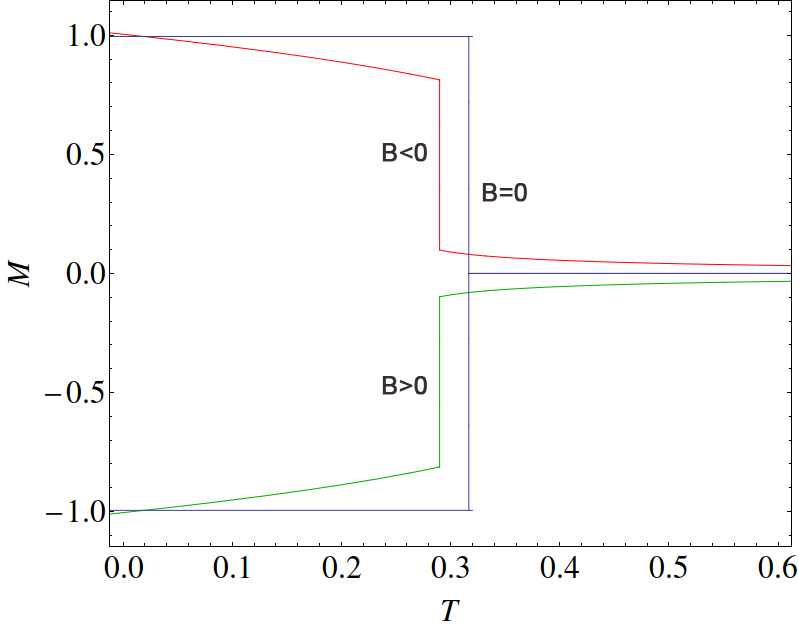}
        \caption{$\M$ vs. $T$ curve for $B\ra 0$ (blue curve) and
          $B\neq 0$ (red and green curve for $B<0$ and $B>0$
          respectively). At the transition temperature a sharp jump
        in magnetization is observed. Low temperature phase has
        constant magnetization and high temperature phase has zero
        magnetization.}\label{fig:MT}
\end{figure}

We study the magnetic properties of a $(2+1)$ dimensional boundary
$CFT$.  The free energy of the $CFT$ is conjectured to be the free
energy of the bulk spacetime. We calculate magnetization of the $CFT$
and find that in $B\ra 0$ limit the boundary theory shows a
ferromagnetic like behaviour. Above a temperature $T_o$ the dominant
phase has zero magnetization whereas below $T_o$ a constant
magnetization phase is dominant (see figure (\ref{fig:MT})). Unlike a
ferromagnetic system though, we find a discontinuity in magnetization
at $T_0$.
In presence of a finite
magnetic field we calculate susceptibility $\chi$ of the system and
observe diamagnetic ($\chi<0$) or paramagnetic ($\chi>0$) behavior of
the system depending on the temperature and magnetic field. We brief
our observation here and present the details in section
\ref{sec:magnetization}.

\begin{itemize}
\item There exists a maximum magnetic field $B^*$ above which the
  thermodynamics is dominated by a diamagnetic phase for any
  temperature. This phase of $CFT$ is dual to the single black hole
  phase in the bulk. See figure (\ref{fig:chiTB1}).
\item For $B_{c}<B<B^*$ ($B_c$ is the critical magnetic field), the
  thermodynamics is again governed by a single phase (which is dual to
  the single black hole phase in bulk). But this phase shows two
  crossovers between diamagnetic and paramagnetic phases. Very high
  and very low temperature phases are diamagnetic, while in between
  the system is paramagnetic. See figure (\ref{fig:chiTB2}).
\begin{figure}[h]
        \begin{subfigure}[b]{0.32\textwidth}
                \centering
                \includegraphics[width=\textwidth]{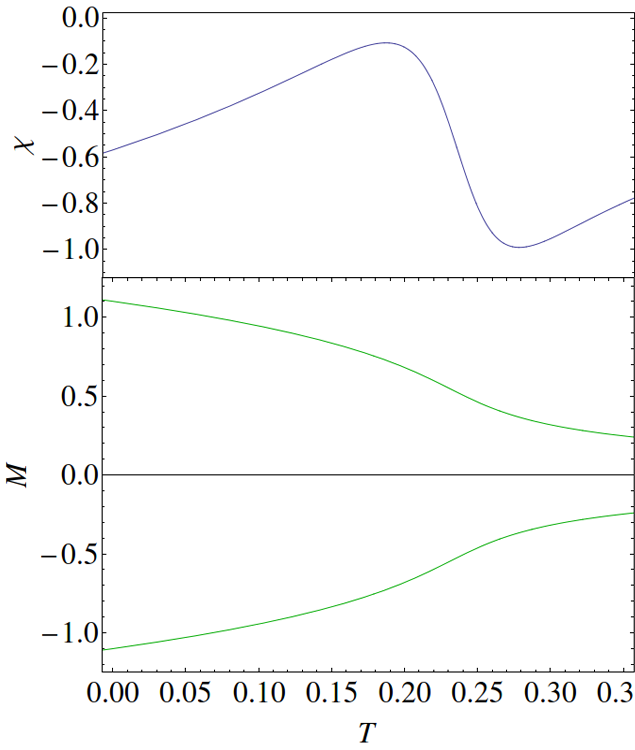}
                \caption{$B>B^*$.}\label{fig:chiTB1}
                 \end{subfigure}
        \begin{subfigure}[b]{0.32\textwidth}
                \centering
                \includegraphics[width=\textwidth]{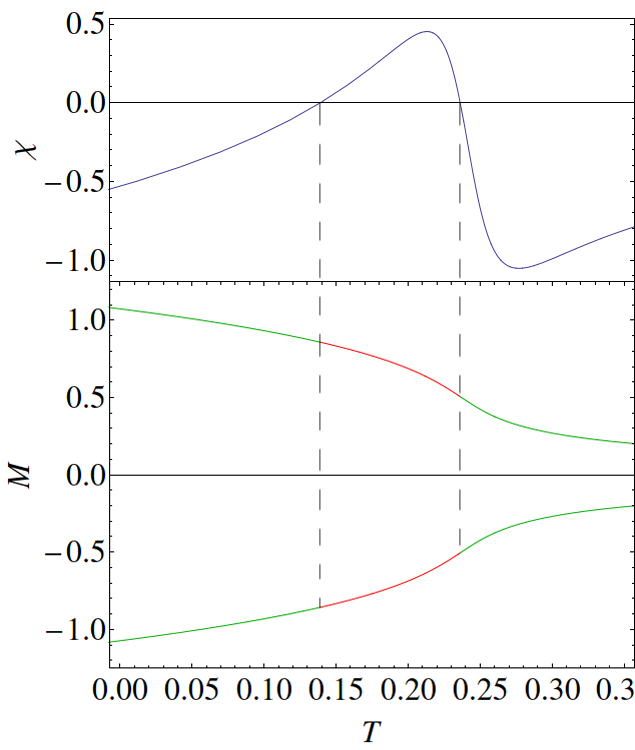}
		\caption{$B^*>B>B_c$. }\label{fig:chiTB2}
         \end{subfigure}
\begin{subfigure}[b]{0.32\textwidth}
                \centering
                \includegraphics[width=\textwidth]{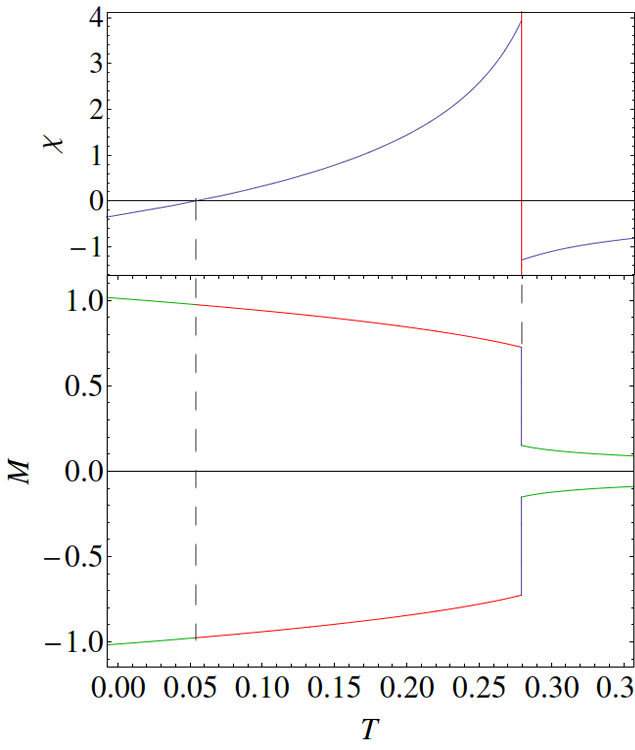}
		\caption{$B$ sufficiently less than $B_c$.}\label{fig:chiTB3}
         \end{subfigure}
         \caption{$\chi-T$ and $\M -T$ plots. Green segments are
           diamagnetic while red are paramagnetic. In plot {\bf (c)}
           the transition from large magnetization phase to small
           magnetization phase is a transition from a paramagnetic
           phase to a diamagnetic phase.}
\end{figure}
\item Criticality appears at $B=B_c$. Two new phases nucleate (one of
  which is thermodynamically unstable). Other two stable phases are
  dual to small and large black holes in bulk. If $B$ is sufficiently
  smaller that $B_c$, we see a transition between a paramagnetic (dual
  to SBH) phase and diamagnetic (dual to LBH) phase at
  $T=T_0$. See figure (\ref{fig:chiTB3}).

\end{itemize}

The organisation of this paper is as follows. In the next section we
give a quick review of dyonic black hole and its holographic
set-up. In section \ref{sec:thermodyna}, we discuss different phases
of dyonic black hole and their thermodynamic properties. Phase
transitions and phase diagrammes have been studied in section
\ref{phasestructure}. Section \ref{sec:magnetization} contains
discussion on magnetic properties of the dual CFT. Stability analysis
has been studied in section \ref{stability}. We conclude our paper
with some remarks in section \ref{conclu_remarks}. There are four
appendices in this paper. In appendix \ref{backgroundsubtraction}, we
provide an alternative method to calculate black hole free energy:
background subtraction method. Appendix \ref{App:MT-BM} contains a
through discussion on critical points in ${\cal M}-B$ plane. We study
symmetry properties of the free energy of the boundary theory in
appendix \ref{App:Ferro}. In the last appendix (appendix
\ref{A:detdiapara}), we provide a detailed study on diamagnetic and
paramagnetic phases of the CFT.

\section{Dyonic black hole solution} \label{dyonic-BH}

$AdS$ dyonic-dilaton black hole solution can be embedded into maximal
gauged supergravity in four dimensions \cite{Lu:2013ura}. However, we
consider a rather simpler solution, which solves the equations of
motion obtained varying the Reissner-Nordstr\"om action in four
dimensions in presence of a cosmological term.

We start with the Reissner-Nordstr\"om action in presence of a
negative cosmological constant,
\begin{equation}
  I=\frac{1}{16 \pi G_4}\int
  \mathrm{d}^4x\sqrt{g}\left(-R+F^2-\frac{6}{b^2}\right) .
\end{equation}

The Equations of Motion are given by,
\be\label{E:EOM1}
 R_{\mu\nu}-\frac{1}{2}g_{\mu\nu}
R-\frac{3}{b^2}g_{\mu\nu} = 2(F_{\mu\lambda}F_\nu^{\
  \lambda}-\frac{1}{4}g_{\mu\nu}F_{\alpha\beta}F^{\alpha\beta}); \ee
\be\label{E:EOM2} \nabla_{\mu}F^{\mu\nu}=0 .
\ee
A static spherically symmetric solution to these equations is given by
\begin{equation}\label{Asol}
	A=\lb-\frac{q_E}{r}+\frac{q_E}{r_+}\rb \df t + \lb q_M \cos\theta \rb 
\df \phi 
\end{equation} 
and 
\be\label{metric}
ds^2=-f(r)\mathrm{d}t^2+\frac{1}{f(r)}\mathrm{d}r^2+r^2\mathrm{d}\theta^2
+ r^2 \mathrm{sin}^2\theta \mathrm{d}\phi^2 
\ee
 where, $A_\mu$ is the
electromagnetic four-potential and 
\be\label{E:themetric} f(r)=\left(
  1+\frac{r^2}{b^2}-\frac{2M}{r}+\frac{q_E^2+q_M^2}{r^2} \right).  
\ee
$q_E$, $q_M$ and $M$ are the integration constants, identified as
electric charge, magnetic charge and mass of the black hole
respectively. $r_+$, the horizon of the black hole is given by
\begin{equation}\label{E:metricf1}
	f(r_+)=\left(
          1+\frac{r_+^2}{b^2}-\frac{2M}{r_+}+\frac{q_E^2+q_M^2}{r_+^2}
        \right)=0 .
\end{equation}
We shall work in an ensemble where the asymptotic value of $A_t$ is
constant, which implies electric potential $\Phi_E$ defined as
\begin{equation}
	\Phi_E=\frac{q_E}{r_+}
\end{equation}
is constant in our thermodynamic analysis.

The Hawking Temperature of this Black Hole is given by
\begin{equation}\label{E:thetemperature}
  T=\frac{1}{\beta}=\frac{1}{4\pi r_+}\left[1+
    \frac{3r_+^2}{b^2}-\Phi_E^2-\frac{q_M^2}{r_+^2} \right].
\end{equation}

\subsection{Holographic dictionary}\label{holodic}

$(3+1)$ dimensional dyonic black hole in asymptotically $AdS$ spacetime
is conjectured to be dual to a $(2+1)$ dimensional $CFT$ living on the
boundary of the $AdS$ space, which has topology $R\times S^2$. The
bulk gauge field is dual to a global $U(1)$ current operator
$J_{\mu}$. The $CFT$ has a conserved global charge $\langle J^t\rangle$
given by
\ben 
\left<
  J^t\right> &=& \frac{q_E}{16 \pi G_4} \nonumber\\
&=& \frac{\sqrt2 N^{3/2} q_E}{24 \pi b^2} 
\een 
where, we use the holographic dictionary
\be
\frac{1}{16 \pi G_4} = \frac{\sqrt2 N^{3/2}}{24 \pi b^2}, \quad N \text{
  is the degree of the gauge group of the $CFT$.}\nonumber
\ee

The boundary $CFT$ also has a constant magnetic field. The strength of
the magnetic field is given by $B = q_M/b^2$ which can be read off
from the asymptotic value of bulk field strength obtained from
equation (\ref{Asol}). We shall study the magnetic properties of this
strongly coupled system using the holographic setup.

In section \ref{sec:magnetization}, we see that the $CFT$ undergoes a
phase transition. A system in the finite volume, in general, does not
exhibit any phase transition. But in the large $N$ limit, {\it i.e.},
when the number of degrees of freedom goes to infinity then it is
possible to have a phase transition even in finite volume.

\section{Thermodynamics of dyonic black hole}
\label{sec:thermodyna}

In our discussion of thermodynamics of dyonic black hole\footnote{For
  thermodynamics of dyonic-dilaton black hole look at
  \cite{Lu:2013ura}.} we identify the cosmological constant $\Lambda
=-6/b^2$ with the thermodynamic pressure of the system. In most
treatments of black hole thermodynamics in $AdS$ space, the
cosmological constant is treated as a constant parameter, $i.e.$
thermodynamics and phase structure of black holes are studied in a
fixed $AdS$ background. In \cite{HenneauxTeitelboim}, the authors
first considered the cosmological constant as a dynamical variable of
the system. Later it was suggested in
\cite{lambdathermo1,lambdathermo2,lambdathermo3,
  lambdathermo4,lambdathermo5,lambdathermo6} that $\Lambda$ can be
treated as a thermodynamic variable of the system. In fact, in
presence of a cosmological constant the first law of black hole
thermodynamics becomes inconsistent with the Smarr relation (the
scaling argument is no longer valid) unless the variation of $\Lambda$
is included in the first law \cite{kastor}. In \cite{kastor} it has
also been shown that once we consider variation of $\Lambda$ in the
first law, the black hole mass $M$ is then identified with enthalpy
rather than internal energy of the system. Therefore, in the following
discussion we take,
\be
P=-\frac{\Lambda}{8\pi} = \frac{3}{8\pi}\frac{1}{b^2}.  
\ee
Thus our ensemble is characterised by $T$, $P$, $q_M$ and
$\Phi_E$. The corresponding free energy $W$ is given
by
\begin{equation}\label{E:freeenergy0}
W=E-TS-\Phi_E q_E .
\end{equation} 

\subsection{On-shell action and background}

We discuss the thermodynamics of black holes in the Euclidean
framework as generally prescribed by Gibbons and Hawking. The
canonical partition function is defined by a functional integral over
metrics with the Euclidean time coordinate $\tau$ identified with
period $\beta$.
\begin{equation}
	\mathcal{Z}=\int [{\cal D}g]\e^{-I_{E}},
\end{equation} 
$I_E$ is the Euclidean action. In the semi-classical limit that we are
considering, the dominant contribution to the path integral comes from
classical solutions to the equations of motion. In this case, 
\be \log
\mathcal{Z} = -I_E^{onshell},  \ee
where $I_E^{onshell}$ is the action evaluated on equations of
motion. Using (\ref{E:EOM1}, \ref{E:EOM2}) we can write
\begin{equation}
  I_E^{onshell}=\frac{1}{16 \pi G_4}\int 
\df^4x\sqrt{g}\left(F^2+\frac{6}{b^2}\right).
\end{equation} 
Free energy is thus given by:
\begin{equation}
	W=-\frac{1}{\beta}\ln\mathcal{Z}=\frac{I_{onshell}}{\beta} .
\end{equation} 

The on-shell action is divergent because the $r$ integration ranges
from $r_+$ to $\infty$. Therefore, we need to regularize the action by
introducing a finite cutoff $\tilde R$. Then, to renormalize the
action, we can either add some counter-term to the original action or
we can subtract the contribution of a background spacetime. Here we
follow the first prescription. Method of background subtraction is
discussed in appendix \ref{backgroundsubtraction}.

For a electromagnetically charged black hole with mass $M$, magnetic
charge $q_M$ and electric potential at infinity $\Phi_E$, the
regularised on-shell action is given by
\begin{align}\label{E:BHonshellaction}
  I_{BH}	&=\frac{1}{16 \pi G_4}\int^{\tilde R}_{r_+}
  d^4x\sqrt{g}\left(F^2+\frac{6}{b^2}\right) \nonumber \\ 
  &=\frac{\beta}{4 G_4}\left[\frac{2(\Phi_E^2r_+^2-q_M^2)}{\tilde
      R}+\frac{6\tilde
      R^3}{3b^2}-\frac{2(\Phi_E^2r_+^2-q_M^2)}{r_+}-\frac{6r_+^3}{3b^
      2} \right] .
\end{align}
Here $\tilde R$ is the cutoff. We shall take $\tilde R\rightarrow
\infty$ at the end. $\beta$ is the period of Euclidean time coordinate
$\tau$, identified as inverse Hawking temperature
\begin{equation}\label{E:HawkingT}
	\beta=\frac{4\pi}{f'(r_+)}=\frac{4\pi
          b^2r_+}{3r_+^2+b^2\left(1-\Phi_E^2-\frac{q_M^2}{r_+^2}\right)}  .
\end{equation} 
Note that the second term in equation (\ref{E:BHonshellaction}) gives
a diverging contribution to free energy as $\tilde R\rightarrow
\infty$. To tame the divergence we add counterterms following
\cite{balakraus}. We find that the on-shell action can be made finite
with the following counterterms,
\begin{equation}
S_{ct} = \frac{1}{8\pi G_4}\int_{\partial {\cal M}} d^3 x
\sqrt{-\gamma} \lb  c_1 + c_2 R^{(3)}\rb. 
\end{equation}
Here $\partial {\cal M}$ is the asymptotic boundary of $AdS$
spacetime, $\gamma$ is the induced metric on the boundary, $R^{(3)}$
is the Ricci scalar calculated for the metric $\gamma$. $c_1$ and
$c_2$ are two numerical constants, their values are given by,
\be
c_1 = -\frac1b , \qquad c_2= \frac{b}{4}.
\ee
Hence, the free energy is given by,
\begin{equation}\label{E:firstfreeenergy}
	W=\frac{I}{\beta}=\frac{1}{4 G_4}\left[ -\Phi_E^2r_+
          +\frac{3q_M^2}{r_+} -\frac{8\pi Pr_+^3}{3} +r_+ \right] . 
\end{equation}

\subsection{Thermodynamic variables and their relations}

Using the expression for free energy $W(T,\Phi_E,q_M,P)$ in equation
(\ref{E:firstfreeenergy}), all the thermodynamical variables can be
computed easily. We quote the results as follows:
\begin{align}\label{thermoquantity}
  q_E &=-\frac{\partial W}{\partial \Phi_E}=\frac{\Phi_Er_+}{G_4} ;
  \nonumber\\
  \Phi_M&=\frac{\partial W}{\partial
    q_M}=\frac{q_M}{G_4\ r_+} ; \nonumber\\
  S &=-\frac{\partial W}{\partial T}=\beta^2\frac{\partial W}{\partial
    \beta}=\frac{1}{4G_4}4\pi
  r_+^2=\frac{A_H}{4G_4} ; \nonumber\\
  V  &=\frac{\partial W}{\partial
    P}=\frac{4\pi}{3}r_+^3 .
\end{align}
Here $\Phi_M$ is the chemical potential corresponding to magnetic
charge $q_M$. Using equation (\ref{E:freeenergy0}) we can calculate
the enthalpy $E$ of the BH given by:
\begin{equation}
	E=W+TS+\Phi_E q_E =\frac{M}{G_4} .
\end{equation} 
Thus we see that enthalpy $E$ is equal to the mass of the black
hole. From now and onwards, we set $G_4=1$.

\subsection{Equation of state}

In \cite{pvcritic} it has been argued that once we consider the
cosmological constant as thermodynamic pressure and the volume covered
by the event horizon as thermodynamic volume, the black hole equation
of state has surprising similarity with equation of state of a Van der
Waals fluid, which describes the liquid-gas phase transition
qualitatively. One can write the pressure as a function of $T$, $r_+$,
$\Phi_E$ and $q_M$ as follows from equation (\ref{E:thetemperature}):
\begin{equation}\label{E:EOS1}
	P=\frac{T}{v}-\frac{1-\Phi_E^2}{2\pi v^2}+\frac{2q_M^2}{\pi v^4}
\end{equation} 
where, $v=2r_+$ can be identified with the specific volume of the
system \cite{pvcritic}. This equation describes different phases of a
dyonic black hole in a fixed electric potential and magnetic charge
ensemble which is similar to extended liquid-gas phase diagram. We
shall discuss the phase structure in section \ref{phasestructure}.

\subsection{Specific heats and thermodynamic stability}

In thermodynamics, heat capacity or thermal capacity is an important
measurable physical quantity. It specifies the amount of heat required
to change the temperature of an object or body by a given
amount. There are two different heat capacities associated with a
system. $C_V :$ measures the heat capacity when the heat is added to
the system keeping the volume constant and $C_P :$ when the heat is
added at constant pressure. Heat capacities can be calculated using
the standard thermodynamic relations:
\begin{equation}
  C_V=T\frac{\partial S}{\partial T}\bigg|_V , \quad\quad
  C_P=T\frac{\partial S}{\partial T}\bigg|_P .
\end{equation} 
Using the expression for black hole entropy (equation
\ref{thermoquantity}) one can calculate the heat capacities:
\begin{equation}\label{E:cvvanish}
  S=\frac{A_H}{4}=\pi r_+^2=\pi\left(\frac{3V}{4\pi}\right)^{2/3}
  \quad \Rightarrow C_V=0 . 
\end{equation} 
Using equation (\ref{E:EOS1}) $C_P$ is given by
\begin{equation}\label{E:cpdiverge}
  C_P=2S\left[\frac{8PS^2 +S(1-\Phi_E^2) - \pi
      q_M^2}{8PS^2-S(1-\Phi_E^2) + 3\pi q_M^2}\right] . 
\end{equation}
 
Thermodynamic stability requires that $C_P>0$. This implies the
stability of different thermodynamic phases are bounded by, 
\be\label{E:stabcp}
\frac13 S(1-\Phi_E^2) -\frac83 PS^2 < \pi q_M^2 < 8P S^2 +
S(1-\Phi_E^2) .  
\ee
We shall get back to this point in section \ref{stability}. 

The specific heat $C_P$ diverges when the denominator vanishes. We
shall discuss this issue in section \ref{criticality} when we discuss
about the critical points in black hole phase diagram.

\section{Black hole phase transition}\label{phasestructure}

$P-V$ diagram for a Van der Waals fluid shows multiple volume
solutions (gas, liquid and a thermodynamically unstable state) for a
given pressure in a specific range. The presence of the unstable state
is unphysical because $\partial P/\partial V|_T $ is positive for this
state. This problem is avoided by Maxwell's equal area law to give the
corrected $P-V$ diagram, which indirectly states that the system will
exist in a phase which has the minimum value of corresponding free
energy. When the free energy of gas and liquid phases crosses each
other, the system instantaneously jumps from the higher free energy
phase to the lower. When the two free energies are equal however, the
gas and liquid phases coexist.

An exactly similar behaviour is observed when we study the equation of
state of a dyonic black hole. Van der Waals equation and black hole's
equation of state are different in nature. The first one is a cubic
equation whereas the second one is quartic. Therefore, black hole's
equation should have four roots (for $v$) for a given set of
parameters. However, it turns out that there exists $maximum$ three
real positive solutions for $v$ (the fourth root is negative in this
case). In this section we study the phase diagram of dyonic black
hole. We find that there is a phase transition between a small black
hole and large black hole which is analogous to the liquid-gas phase
transition. We also discuss the limit of this phase transition as
$q_M\rightarrow 0$ and show that it reduces to the Hawking-Page phase
transition \cite{hawkingpage}.

\subsection{Different phases of black holes} \label{phasesofBH}

We write $P$ as a function of thermodynamic volume $V$ from equation
(\ref{E:EOS1}) 
\be\label{E:EOS2}
P=\frac{T}{V^{1/3}}\lb\frac{\pi}{6}\rb^{1/3}-\frac{1-\Phi_E^2}{2\pi
  V^{2/3}}\lb\frac{\pi}{6}\rb^{2/3}+\frac{2q_M^2}{\pi
  V^{4/3}}\lb\frac{\pi}{6}\rb^{4/3} .  \ee

\begin{figure}[h]
        \begin{subfigure}[b]{0.5\textwidth}
                \centering
                \includegraphics[width=\textwidth]{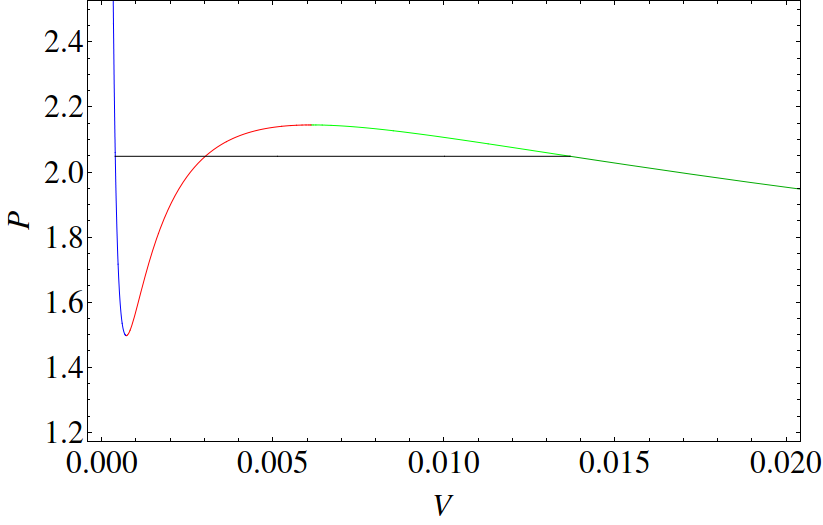}
		\caption{}\label{fig:PV}
        \end{subfigure}
	\begin{subfigure}[b]{0.5\textwidth}
                \centering
                \includegraphics[width=\textwidth]{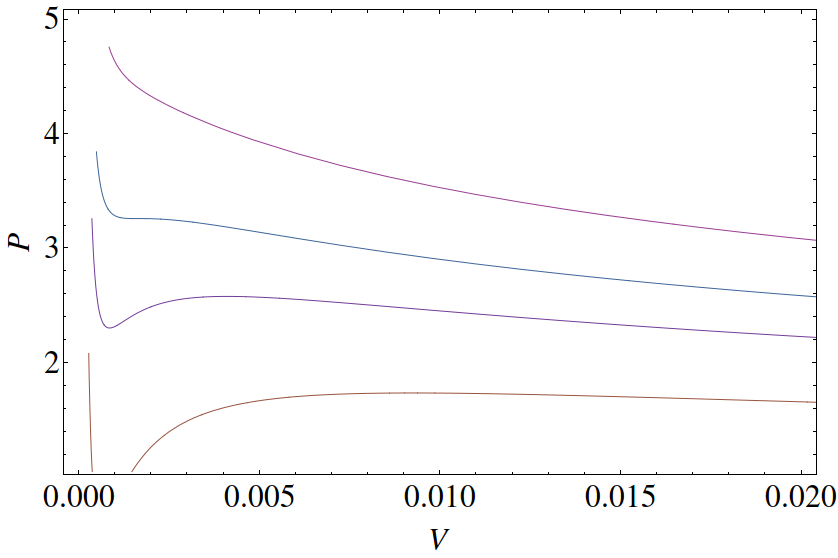}
		\caption{}\label{fig:tpv}
        \end{subfigure}
	\caption{\textbf{(a)} $P-V$ diagram for all other parameters
          held constant ($q_M=0.028$, $\Phi_E=0.35$, $T=1.05$). The
          black line refers to the coexistence pressure. Blue, red and
          green parts of the curve corresponds to Branch-1,2,3
          respectively. \textbf{(b)} $P-V$ diagram for fixed
          $q_M(=0.028)$ and $\Phi_E(=0.35)$ and varying $T$. The
          second curve from the top corresponds to the Critical
          Point.}
\end{figure}

The $P-V$ plot (figure \ref{fig:PV}) we get keeping $T$, $\Phi_E$ and
$q_M$ fixed, are identical to that of liquid-gas case in Van der Waals
case. In figure (\ref{fig:tpv}) we plot the same curves for varying
$T$.

In the first figure we see that for a particular value of $P$ in a
window, we have three different solutions for black hole
horizon. These are identified as Branch-1 (Blue), Branch-2 (Red) and
Branch-3 (Green) (figure \ref{fig:PV}). For large $P$ only Branch-1
(SBH) exists, while for low $P$ Branch-3 (LBH) is the only
solution. As we shall see later, Branch-2 always gives a
thermodynamically unstable phase. The remarkable resemblance with the
liquid-gas phase transition in Van der Waals case is clearly apparent
here.

As we alter the Temperature (see figure \ref{fig:tpv}), we notice that
above a critical temperature, Branch-2 totally disappears and Branch-1
and 3 coalesce. Below the critical temperature however, there exist 3
phases of black hole in a particular pressure window, out of which
one is unstable.

\begin{figure}[h]
        \begin{subfigure}[b]{0.5\textwidth}
                \centering
                \includegraphics[width=\textwidth]{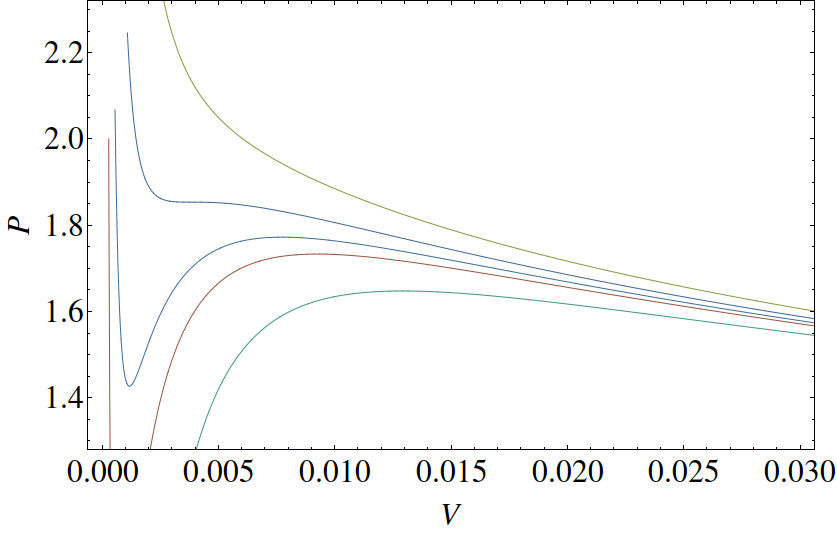}
		\caption{}\label{fig:qpv}
        \end{subfigure}
	\begin{subfigure}[b]{0.5\textwidth}
                \centering
                \includegraphics[width=\textwidth]{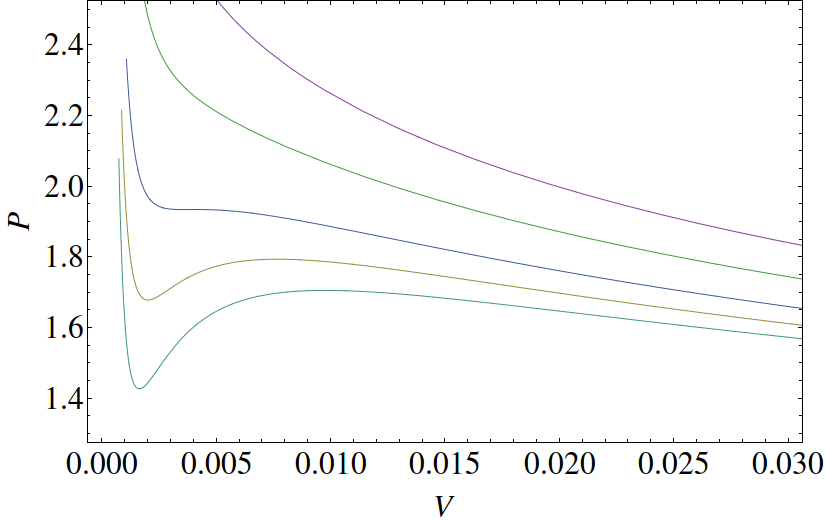}
		\caption{}\label{fig:ppv}
        \end{subfigure}
	\caption{\textbf{(a)} $P-V$ diagram for fix $T(=0.96)$ and
          $\Phi_E(=0.35)$ and varying $q_M$. The last curve
          corresponds to $q_M=0$. \textbf{(b)} $P-V$ diagram for fix
          $q_M(=0.038)$ and $T(\approx1)$ and varying $\Phi_E$. The
          last curve corresponds to $\Phi_E=0$.}
\end{figure}

Unlike Van der Waals equation, we have 2 more free parameters apart
from $T$: $q_M$ and $\Phi_E$. As we vary $q_M$ the behavior of $P-V$
diagram is quite similar to what we have seen for varying $T$ (See
figure \ref{fig:qpv}). Here also, there exists a critical magnetic
charge $q_{M(c)}$ above which there exists only one branch of
solution. Nucleation of other two branches starts at $q_{M(c)}$
(depending on the values of other parameters). \\

\noindent
{\bf \underline {$q_M \ra 0$ limit}: }\ \ \ 
We note an interesting and important behavior as $q_M\rightarrow
0$. Branch-1 overlaps with y-axis i.e. $r_+$ for small black hole
approaches to zero in $q_M\ra 0$ limit. Thus the small black hole
solution evaporates to a global $AdS$, which exists for all values of
pressure. However, the other two branches still persist in this limit.

Finally, we vary $\Phi_E$ keeping $T$ and $q_M$ fixed (figure
\ref{fig:ppv}). From the equation of state we can easily check that if
$\Phi_E^2>1$ then there exists only one branch for any values of $T$
and $q_M$. Criticality starts at $\Phi_E^2=1$ and persists for $0<
\Phi^2_E<1$. At $\Phi_E=0$, which is a purely magnetically charged
system, the phase diagram is similar\footnote{Since the metric
  (\ref{E:themetric}) is symmetric in electric and magnetic charges,
  pure constant electric charge system behaves similar to the pure
  constant magnetic charge ensemble.} to electrically charged black
hole in fixed charge ensemble of \cite{myers,pvcritic}.

\subsection{Critical points}\label{criticality}

Studying the figure (\ref{fig:tpv}), (\ref{fig:qpv}) and
(\ref{fig:ppv}) in the last section we see that there exists a
critical hyper-surface in the parameter space ($T,\Phi_E$ and $q_M$)
where Branch-1 and 3 coalesce and Branch-2 vanishes. If we go beyond
that hyper-surface we have only one branch and below that
hyper-surface we have three different branches which is much alike the
critical point in liquid-gas phase transitions. But the difference is,
in case of liquid-gas phase diagram the critical point was determined
by the temperature only, whereas in this case it is a two dimensional
hyper-surface. The critical surface can be obtained in the following
way.

For multiple phases to occur in our system, we need a region of
positive slope in $P-V$ diagram. Using equation (\ref{E:EOS1}) we thus
have the bound:
\begin{equation}\label{E:physicalcond}
	-\frac{\pi}{2} v^5
        P'=Tv^3-\frac{1-\Phi_E^2}{\pi}v^2+\frac{8q_M^2}{\pi}<0 .
\end{equation} 
Real positive solution for $v$ is possible when
\begin{align}
	|\Phi_E|< 1 \quad \text{and} \quad (1-\Phi_E^2)^3-54\pi^2q_M^2T^2\geq 0 .
\end{align} 
The equality holds at the critical point where Branch-2 just vanishes,
giving us the equation for the critical surface:
\begin{equation}\label{E:thirdcritical}
	3\sqrt{6}\pi q_{M(c)} T_c=(1-\Phi_{E(c)}^2)^{3/2}.
\end{equation}
Using the equation of state we find the expression for critical
pressure and volume as
\begin{equation}\label{E:forthcritical}
  P_c=\frac{9\pi T_c^2}{16 \lb 1-\Phi_{E(c)}^2 \rb} , \qquad
  v_c=\frac{2\sqrt{6}q_{M(c)}}{\lb1-\Phi_{E(c)}^2  \rb^{1/2}} . 
\end{equation} 
The relations (\ref{E:thirdcritical}) and (\ref{E:forthcritical}) can
also be reached directly by demanding divergence of $C_P$ in equation
(\ref{E:cpdiverge}). This designates the boundary of multiple-phase
regime.

\subsubsection{Critical exponents}

On the critical surface given by equation (\ref{E:thirdcritical}) and
(\ref{E:forthcritical}), we calculate the ratio of critical pressure
times critical volume over critical temperature. The result is given
by
\begin{equation}
	\frac{P_cv_c}{T_c}=\frac{3}{8},
\end{equation} 
which is a universal constant and is exactly the same as for the Van
der Waals fluid. Defining new reduced variables
\begin{equation}
  p_R=\frac{P}{P_c} \quad\quad t_R=\frac{T}{T_c} \quad\quad v_R=\frac{v}{v_c},
\end{equation} 
equation of state (equation \ref{E:EOS1}) can be written as (which has
the same form as in \cite{pvcritic})
\begin{equation}\label{E:EOS3}
	8t_R=3v_R\lb p_R+\frac{2}{v_R^2}\rb-\frac{1}{v_R^3} .
\end{equation} 

We also calculate the critical exponents defined as, \ben
\text{Specific heat}: && C \sim |T-T_c|^{-\alpha}\\ \nonumber
\text{order parameter}: && {\cal O} \sim |T-T_c|^{\beta} \\ \nonumber
\text{susceptibility or compressibility}: && \chi (\text{or} \ \
\kappa) \sim |T-T_c|^{-\gamma} \\ \nonumber \text{equation of state}:
&& (p-p_c) \sim (v-v_c)^{\delta}.  \een In this case the order
parameter is given by the difference between the radii of large and
small black hole ($|r_+^{L}-r_+^{S}|$). Calculation of these critical
exponents is fairly straight forward. Our final results are given by,
\begin{equation}
  \alpha=0 , \quad\quad \beta=\frac{1}{2} ,\quad\quad \gamma=1
  ,\quad\quad   \delta=3 .
\end{equation} 
Note that these results match with mean-field theory.

\subsection{Phase transition}\label{PhaseTransition}

\begin{figure}[h]
        \begin{subfigure}[b]{0.5\textwidth}
          \centering
                \includegraphics[width=\textwidth]{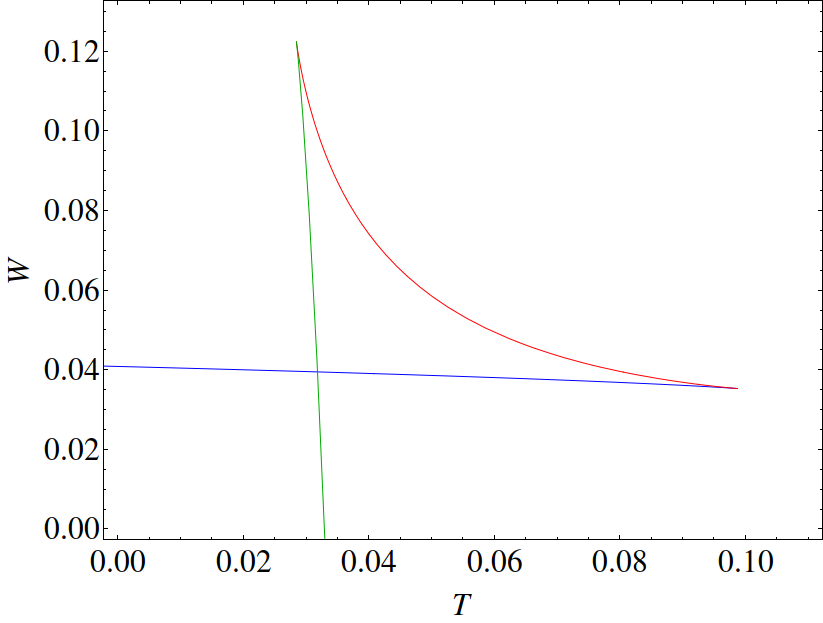}
		\caption{}\label{fig:WT}
        \end{subfigure}
	\begin{subfigure}[b]{0.5\textwidth}
                \centering
                \includegraphics[width=\textwidth]{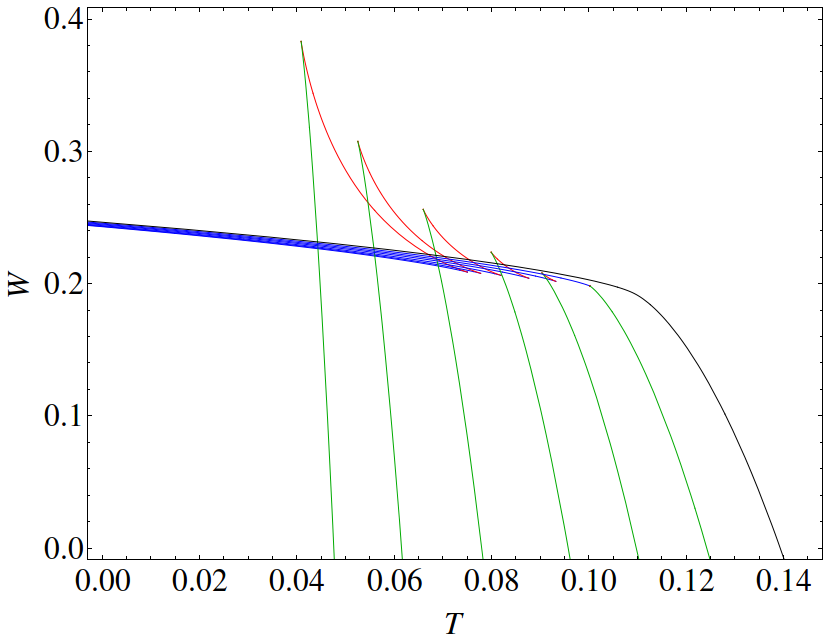}
		\caption{}\label{fig:WTP}
        \end{subfigure}
	\caption{\textbf{(a)} $W-T$ plot for all other parameters
          fixed ($P=0.0035$, $\Phi_E=0.8$, $q_M=0.068$). The graph has
          three separate branches: Branch-1 (Blue), Branch-2 (Red) and
          Branch-3 (Green). \textbf{(b)} $W-T$ diagram for fixed
          $\Phi_E=0.5$, $q_M=0.28$ and varying $P$. $P\approx 0.018$
          is the critical point in this case, after which only one
          solution of BH remains (the black curve).}
\end{figure}

In order to study phase transition we look at the free energy as a
function of temperature for constant $P$, $q_M$ and $\Phi_E$. For
better understanding, we use different colors for three different
branches that appear in $P$-$V$ diagram. In figure (\ref{fig:WT}), the
blue, red and green lines are the free energies for small, unstable
and large black hole as a function of temperature (same as in figure
\ref{fig:PV}). From this plot it is clear that at low temperature
there exists only one branch. As we increase temperature two new
branches appear (the upper cusp) at $T=T_c$ (black hole nucleation
temperature). At this temperature the free energies of two new
branches are greater than the first one, hence the first branch
dominates the thermodynamics. Upon further increase of temperature, we
see that at some temperature ($T=T_o$) the green curve crosses the
blue one and after that the free energy of the large black hole
dominates over the other two branches. This implies a phase transition
between small black hole and large black hole at $T_o$. There exists a
temperature $T_A>T_o$, where Branch-1 and 2 vanishes, is called
annihilation temperature ($T_A$).

The unstable black hole (branch-2, the red curve in the figure \ref{fig:WT}) is
always thermodynamically disfavored as the free energy of this branch
is greater than the free energy of the other two branches for any
temperature. Branch-1 and Branch-3 intersects at the coexistence point
$T_o$ ($\approx 0.03$) which is similar to liquid gas phase
transition.

\begin{figure}[h]
        \begin{subfigure}[b]{0.5\textwidth}
                \centering
                \includegraphics[width=\textwidth]{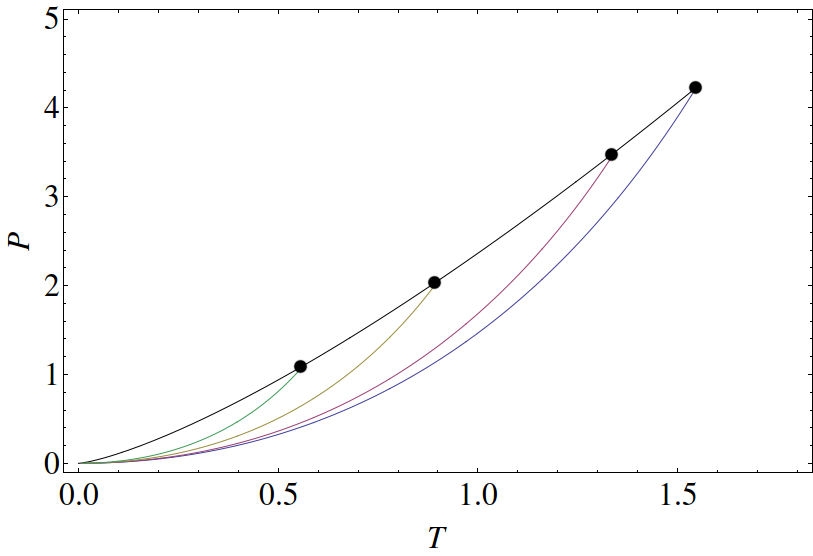}
		\caption{}\label{fig:coexTP}
        \end{subfigure}
	\begin{subfigure}[b]{0.5\textwidth}
                \centering
                \includegraphics[width=\textwidth]{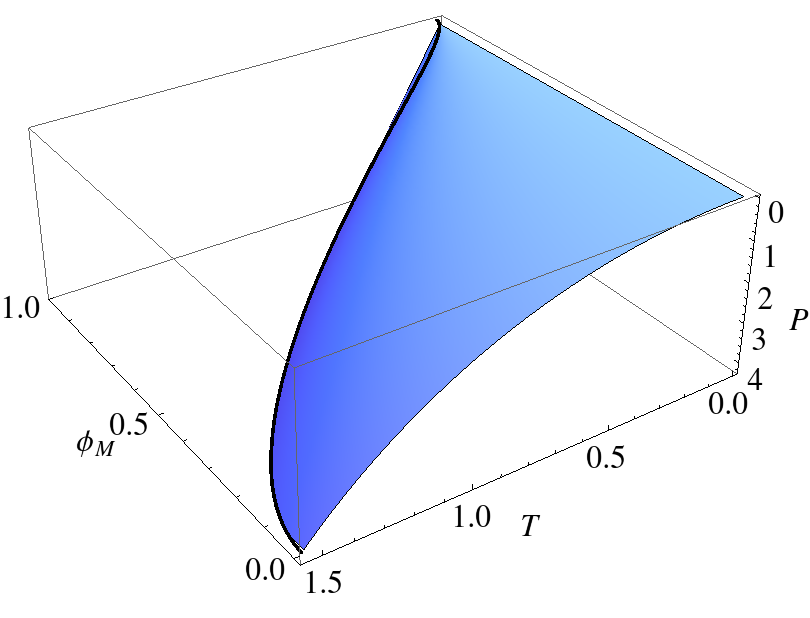}
		\caption{}\label{fig:coexall}
        \end{subfigure}
	\caption{\textbf{(a)} Colored curves are the Coexistence
          Curves in $P-T$ phase space for different values of
          $\Phi_E$. The rightmost is for the lowest value of $\Phi_E$
          which is 0. The black curve is however the critical curve
          for arbitrary $\Phi_E$. \textbf{(b)} Coexistence Plane in
          $P-T-\Phi_E$ phase space. Above the plane lies the Small BH,
          while below the plane we have Large BH. The Black Graph is
          the Critical Curve.}
\end{figure}

The coexistence point can be reached easily by demanding free energy
($W$) to be same for two different volumes: $V_1 (r_+^{(1)})$ and $V_2
(r_+^{(2)})$ which are the roots of equation (\ref{E:EOS2}) for some
$P, T, q_M, \Phi_E$ such that $P'(V)<0$. As we vary the temperature,
black hole never goes to Branch-2 phase, but jumps from Branch-1 to
branch-3 directly at the coexistence point. Therefore, we correct the
$P-V$ diagram by replacing the oscillating part by a flat line (See
figure \ref{fig:PV}). The flat line can be obtained at some particular
pressure called coexistence pressure $P_{co}$ where the free energy
$W$ of both the phases (1 and 3) are same. This is the Maxwell
construction. Needless to mention, the coexistence pressure depends on
the temperature (also on other two parameters in this case $q_M$ and
$\Phi_E$). In figure (\ref{fig:coexTP}) we plot coexistence pressure
vs. temperature for different values of $\Phi_E$.  Coexistence points
form a plane in $T-P-\Phi_E$ plane (keeping $q_M$ fixed).

\subsubsection{Hawking-Page phase transition}\label{hptrans}

\begin{figure}[h]
        \centering
    	\includegraphics[width=0.5\textwidth]{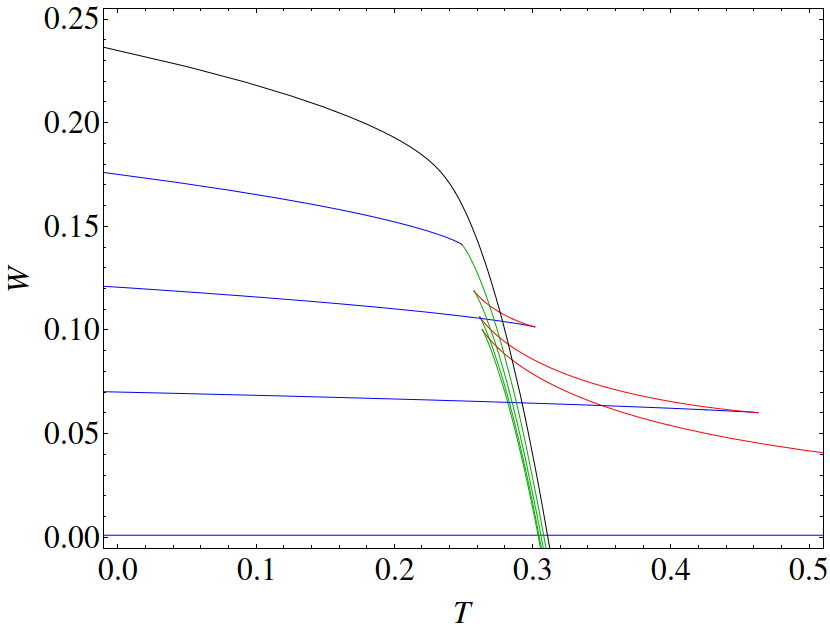}
	\caption{W vs. T for fixed $P=0.12$, $\Phi_E=0.045$ and
          varying $q_M$. Criticality occurs at $q_M=0.173$. The
          bottommost graph is for $q_M=0$.}\label{fig:q0}
\end{figure}

In figure (\ref{fig:q0}) we plot free energy vs. temperature for
different values of $q_M$, keeping $\Phi_M$ fixed. In this section we
investigate the $q_M\ra 0$ limit carefully. From the plot it is clear
that for large $q_M$ we can not see any phase transition (the topmost
plot). There exists only one stable black hole. At $q_M=q_{M(c)}$ two
new branches appear one of them is unstable. As we further decrease
$q_M (>0)$ the $W-T$ plot is same as in figure (\ref{fig:WT}), what we
have already discussed. Here we discuss the $q_M\ra 0$ limit of these
plots (the bottom most curve). In this limit we see that the blue line
(corresponding to small black hole) overlaps with x-axis. This implies
that the free energy of the small black hole, in this particular
limit, reduces to zero. That is, the small black hole reduces to a
global $AdS$ spacetime (the size of this black hole also reduces to
zero in this limit, as we have discussed in section
\ref{phasesofBH}). Hence, the LBH - SBH phase transition reduces to
Hawking-Page phase transition (a transition between black hole and a
global $AdS$ spacetime).

An important point to note here is that, in $q_M\ra0$ limit $r_+$ also
goes to zero but the ratio $q_M/r_+$ remains fixed (which we call
$\Phi_M$). Therefore, the global $AdS$ spacetime has a constant
$\Phi_M$. Since we are working in a constant $\Phi_E$ ensemble, the
global $AdS$ space has a constant electric potential as well.

This completes our discussion on black hole phase transition in
presence of an arbitrary magnetic charge. We have established the
resemblance of Small-Large black hole phase transitions in
asymptotically $AdS$ space (in constant electric potential and magnetic
charge ensemble) with the liquid-gas phase transition.

\subsubsection{Planar black hole}

Planar black holes have horizon topology $R\times R^2$. Black holes
with $R^2$ horizon topology (space part only) can arise in
asymptotically $AdS$ space. These black hole are also called {\it
  Ricci flat} black holes. In our analysis, if we consider the black
hole mass (or $r_+$) to be very large, all the equations and
expressions reduces to those of planar black hole.

Phase diagram for planar black hole does not show any kind of
criticality. For any value of $\Phi_E$ and $q_M$ there exists only one
branch for any temperature. This is clear from the equation of
state for planer black hole,
\begin{equation}\label{E:EOSplaner}
	P=\frac{T}{v}+\frac{\Phi_E^2}{2\pi v^2}+\frac{2q_M^2}{\pi v^4} .
\end{equation} 
Because of the
$Ricci\ flat$ horizon topology planar black hole is always
thermodynamically stable for any temperature. However, planar black
hole with toroidal horizon topology undergoes a phase transition from
black hole to $AdS$ soliton phase \cite{pbh1, pbh2}. It would be
interesting to check if magnetic charge has any nontrivial effect on
this phase transition.

\section{Magnetic properties of boundary
  CFT} \label{sec:magnetization}

We have discussed in section \ref{holodic} that the boundary CFT has a
constant magnetic field due to presence of a magnetic charge in the
bulk spacetime. The magnetic field is given by,
\begin{equation}
	B=\frac{q_M}{b^2} .
\end{equation} 
When we consider thermodynamics of the boundary theory we define a
variable $\cal M$ (magnetization), conjugate to the external magnetic
field $B$. Different phases of boundary theory are also characterised
by this new variable $\M$ defined by the following
relation\footnote{Since we are studying the thermodynamic properties
  of a boundary theory in the context of the $AdS/CFT$ we keep the
  radius of the $AdS$ space to be constant.},
\be 
\M = - \frac{\partial W}{\partial B}\bigg |_{T} .  
\ee
Using the expression (\ref{E:firstfreeenergy}) for $W$ we find,
\be\label{Mvalue}
\M = - \frac{\partial W}{\partial B}\bigg |_{T} = -b^2 \lb\frac{q_M}{r_+} \rb .
\ee
It is worth to note that the positive definiteness of $r_+$ implies
$\mathcal{M}$ and $B$ always have opposite sign.

We calculate free energy and temperature of the $CFT$ in terms of $B$
and $\M$ (boundary parameters) to study the phase structure of the
system, 
\ben W&=&\frac{1}{4}\left[
  -(1-\Phi_E^2)b^4\frac{B}{\mathcal{M}} - 3B\mathcal{M}
  +b^{10}\frac{B^3}{\mathcal{M}^3} 
\right], \label{E:freeenergyBM} \\
T &=& \frac{1}{\beta}=\frac{1}{4\pi}\left[-(1-\Phi_E^2)\frac{1}{b^4}
  \frac{\mathcal{M}}{B}
  -3b^2\frac{B}{\mathcal{M}}+\frac{1}{b^{8}}\frac{\mathcal{M}^3}{B}
\right] \label{E:tempBM} .  
\een

\subsection{Magnetization of boundary phases}

In this section we discuss different phases of the boundary $CFT$ dual
to different black hole phases. W in equation (\ref{E:freeenergyBM})
can be plotted against $T$ for varying $B$. Since $B$ is proportional
to $q_M$, the plot is the same as in figure (\ref{fig:q0}).

\begin{figure}[h]
        %\begin{subfigure}[b]{0.5\textwidth}
                \centering
                \includegraphics[width=0.5\textwidth]{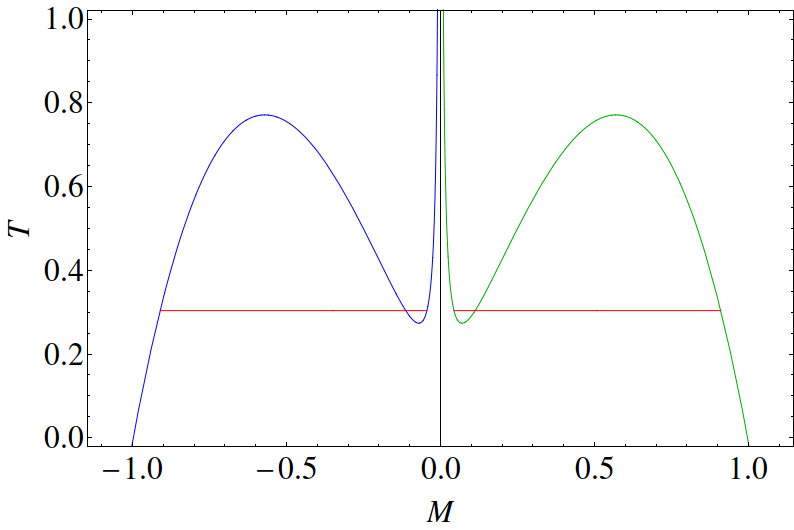}
         %       \caption{}\label{fig:cortm}
        %         \end{subfigure}
                \caption{$T-\mathcal{M}$ plot and Maxwell's
                  construction. Red line is the co-existence point
                  (corresponds to temperature $T_0$). }\label{fig:cortm}
\end{figure}

For $B$ above a critical value $B_c$ (which is proportional to
$q_{M(c)}$, discussed in Section \ref{phasesofBH}) there exists only
one phase of boundary $CFT$ and magnetization of this phase is a
continuous function of temperature.  As $B$ goes below $B_c$ the
boundary theory develops two more new phases. Therefore, in this case
the theory has three different phases (in a given temperature
window). See figure (\ref{fig:cortm}).

The blue (green) curve in figure (\ref{fig:cortm}) corresponds to
positive (negative) magnetic field. In this figure we see that, below
a critical temperature there exists only one phase with high
magnetization (dual to small black hole). Above the critical
temperature, there are three possible phases with different
magnetization. Among them the middle one is thermodynamically
unstable. The phase with small magnetization correspond to the large
black hole phase in the dual theory. The red line indicates the
transition temperature $T_o$. Above this temperature, thermodynamics
is dominated by the phase with small magnetization. Therefore a sharp
jump in magnetization is observed at the transition temperature. Later
in section \ref{suscep}, we discuss about this transition in
detail. In figure (\ref{fig:MT}) we plot the same graph removing the
unstable branch using Maxwell's construction. $\M$ vs. $T$ plots for
different values of magnetic field and $B$ vs. $\M$ plots for
different values of temperature are discussed in Appendix
\ref{App:MT-BM}.

\begin{figure}[h]
	\centering
	\includegraphics[width=0.5\textwidth]{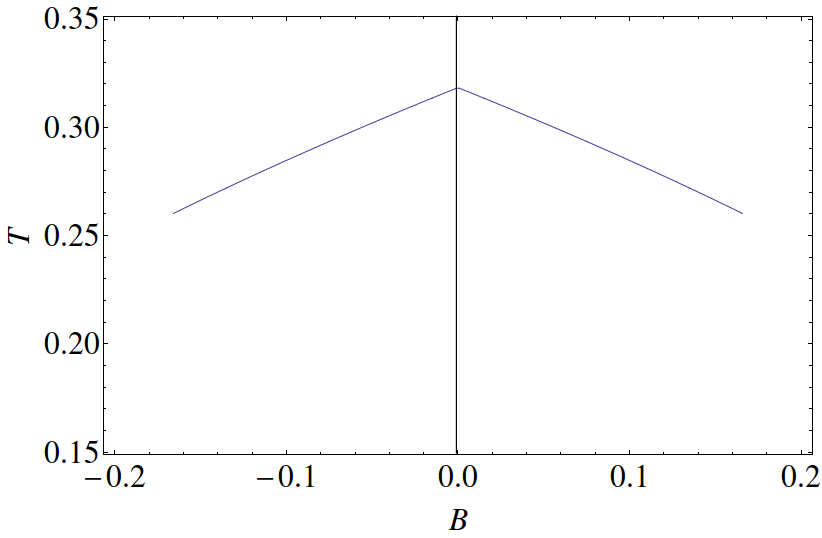}
        \caption{$T_0$ vs. $B$ curve (for $\Phi_E=0.8$). The end point
          is the critical point (for a fixed $\Phi_E$). Above (below)
          the line small (large) magnetization black hole dominates
          the thermodynamics.}\label{fig:B-Tplot}
\end{figure}

The transition temperature $T_o$ depends on the external magnetic
field. As we increase the magnetic field from 0 up to $B_c$ the
transition temperature decreases (figure \ref{fig:B-Tplot}). For $B>B_c$
there exists only one stable branch.

\subsubsection{$B \ra 0$ limit: Ferromagnetic like behaviour}

$B\ra 0$ limit, in particular is interesting. In this limit we see
that the low temperature phase has a constant magnetization whereas
the high temperature phase has zero magnetization. There is a
transition between zero magnetization phase to constant magnetization
phase as we decrease the temperature (See figure \ref{fig:MT}). Unlike
ferromagnetic materials, here we find that the magnetization is
discontinuous at the transition temperature. The constant
magnetization phase is dual to global $AdS$ phase in the bulk. As we
have discussed before, in this limit the radius of small black hole
goes to zero with $q_M/r_+$ fixed. In other words the small black hole
evaporates to global $AdS$ with constant $\Phi_E$ and $\M$. A more
detailed discussion can be found in Appendix \ref{App:Ferro}.

As we have explained before, in the limit $B\ra 0$ LBH/SBH phase
transition reduces to Hawking-Page phase transition. Therefore, low
temperature phase (dual to global $AdS$) of the boundary theory has
zero free energy whereas the high temperature phase dual to LBH has
free energy of order $N^{3/2}$ in the limit $N\ra \infty$. This phase
transition is identified with confinement-deconfinement phase
transition of gauge theory \cite{Witten:1998zw}. Therefore, we see
that the confined phase has a constant magnetization whereas the
deconfined phase has zero magnetization.

\subsection{Magnetic susceptibility of boundary theory} \label{suscep}

Depending on the Temperature and applied Magnetic Field, boundary
theory shows either diamagnetic or paramagnetic behavior. To study the
same we calculate magnetic susceptibility using the formula:
\begin{equation}
	\chi=\left. \frac{\partial \mathcal{M}}{\partial
            B}\right\vert_T .
\end{equation}

A system is said to be diamagnetic if $\chi<0$ and paramagnetic if
$\chi>0$. The magnetic properties of a physical substance mainly
depend on the electrons in the substance. The electrons are either
free or bound to atoms. When we apply an external magnetic field these
electrons react against that field. In general one can see two
important effects. One, the electrons start moving in a quantised
orbit in presence of the magnetic field. Two, the spins of the
electrons tend to align parallel to the magnetic field. One can
neglect the effect of atomic nuclei compared to these two effects, as
they are much heavier than electrons. The orbital motion of electrons
is responsible for diamagnetism whereas, alignment of electrons' spin
along the magnetic field gives rise to paramagnetism. In a physical
substance, these two effects compete. In diamagnetic material the
first one (orbital motion) is stronger than the second one, and
vice-versa in a paramagnetic material.

The temperature in (equation \ref{E:tempBM}) can be written in the
following differential form for a constant $\Phi_E$:
\begin{equation}
  \df T = \frac{\partial T}{\partial B}\df B + \frac{\partial
    T}{\partial \mathcal{M}}\df \mathcal{M} .
\end{equation} 
$\chi$ from this expression can be written as
\begin{equation}\label{E:chiB}
  \chi=\left. \frac{\df \mathcal{M}}{\df B}\right\vert_T =
  \frac{\partial T}{\partial B}\frac{\partial \mathcal{M}}{\partial
    T}=\frac{\mathcal{M}}{B}\lb\frac{3b^{10} B^2+\mathcal{M}^4 -
    b^4\mathcal{M}^2\left(1-\Phi_E^2\right)}{3b^{10}
    B^2+3\mathcal{M}^4 - b^4\mathcal{M}^2\left(1-\Phi_E^2\right)}\rb .
\end{equation} 
We plot $\chi$ against $T$ for various $B$ to study the magnetic
behaviour of the system. Our results are as follows:

\begin{enumerate}
\item There exists a magnetic field $B^*> B_c$, above which the $CFT$
  is diamagnetic for all temperatures (see figure \ref{fig:chiTB1}).

\item For $B_c<B<B^*$, still the boundary theory has a single phase
  but this phase shows two crossovers between paramagnetic and
  diamagnetic phases. At high and low temperature the system behaves
  like a diamagnetic system, while in between it shows paramagnetic
  behavior. See figure (\ref{fig:chiTB2}).

\item For $B<B_c$ (but still close to $B_c$), the unstable branch of
  BH pops up. Thus we see a phase transition at $T_o$ from Small BH
  branch to Large BH branch, both of them being paramagnetic. As
  temperature is decreased (increased), Small BH (Large BH) branch
  crosses over to a diamagnetic phase. Figure (\ref{fig:chiTB4}) shows
  the segment of the curves near $T=T_o$.
\item
\begin{figure}[h]
                \centering
                \includegraphics[width=0.5\textwidth]{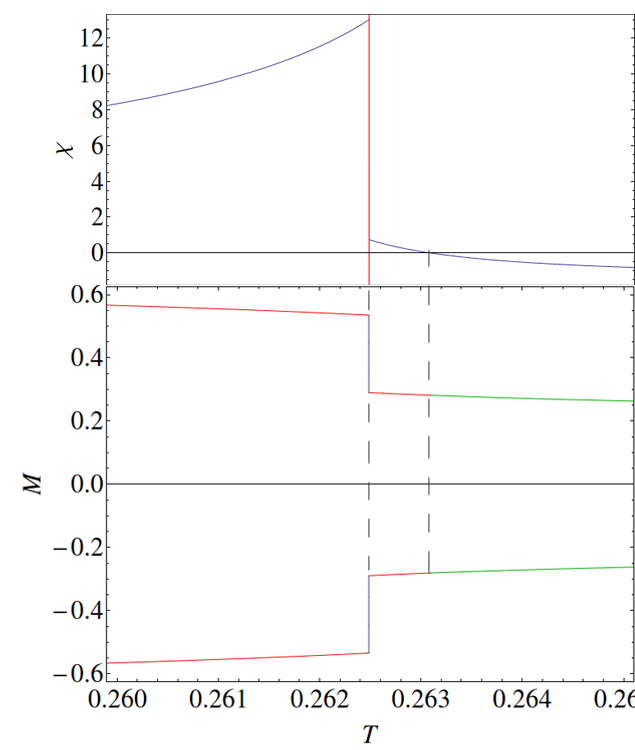}
                \caption{$\M$ vs. $T$ and corresponding $\chi$ vs. $T$
                for $B<B_c$ (but close to it). In this plot it is
                observed that small magnetization branch has $\chi>0$
                for temperature close to the transition
                temperature. As we increase temperature, it crosses
                over to diamagnetic phase $\chi<0$. Similarly, low
                temperature phase near the transition temperature hase
              $\chi>0$ but as temperature is decreased its
              susceptibility becomes negative (the plot does not show
              the low temperature behaviour.)}\label{fig:chiTB4}
\end{figure}
Below $B_c$, when magnetic field is even below a certain value $ B^\#
$ (discussed in Appendix \ref{A:detdiapara}), the paramagnetic branch
of Large BH gets cut off in the Maxwells' construction (figure
\ref{fig:chiTB3}). Thus the phase transition occurs from paramagnetic
Small BH to diamagnetic Large BH.

\end{enumerate}

\subsubsection{High temperature behavior of magnetic susceptibility}

In the high temperature limit, as we know that only the small
magnetization solution (Large BH) dominates, temperature (equation
\ref{E:tempBM}) has the leading contribution from:
\begin{equation}
	T\approx-\frac{3 B}{4 \pi  \mathcal{M}}
\end{equation} 

and $\chi$ for large $T$ will thus be given by:
\begin{equation}
	\chi\approx \frac{\mathcal{M}}{B}\approx\lb-\frac{3}{4\pi}\rb 
\frac{1}{T} .
\end{equation} 

Hence we see that the Curie's Law is satisfied with a negative Curie
Constant $-3/4\pi$.

\section{Stability analysis}\label{stability}

We conclude our discussion by analysing the stability of black hole
solutions. When entropy $S$ is a smooth function of extensive
variables $x_i$'s then sub-additivity of entropy is equivalent to the
Hessian matrix $\ltb \frac{\pa^2 S}{\pa x_i\pa x_j}\rtb$ being
negative definite. For canonical ensemble the only extensive variable
is mass (or energy), therefore, sub-additivity of entropy implies that
$C_P>0$.  For grand canonical ensemble, the variables are mass and
charges therefore the stability lines are determined by finding the
zeros of the determinant of the Hessian matrix. It has been argued in
\cite{gubser} that the zeros of the determinant of the Hessian of $S$
with respect to $M$ and $q_i$'s coincide with the zeros of the
determinant of the Hessian of the Gibbs (Euclidean) action,
\be
I_G = \beta\lb M -\sum_i q^{\text{phys}}_i \Phi_i\rb -S 
\ee
with respective to $r_+$ and $q_i$'s keeping $\beta$ and $\Phi_i$'s
fixed. Note that $q_i$'s are the charge parameters entering into the
black hole solutions where $q_i^{\text{phys}}$'s are the physical
charges. Though this criteria can figure out the instability line in
the phase diagram but it is unable to tell which sides of the phase
transition lines correspond to local stability. One can figure out the
stability region by knowing the fact that zero chemical potential and
high temperature must correspond to a stable black hole solution.

\begin{figure}[h]
	\centering
        \begin{subfigure}[b]{0.4\textwidth}
                \centering
                \includegraphics[width=\textwidth]{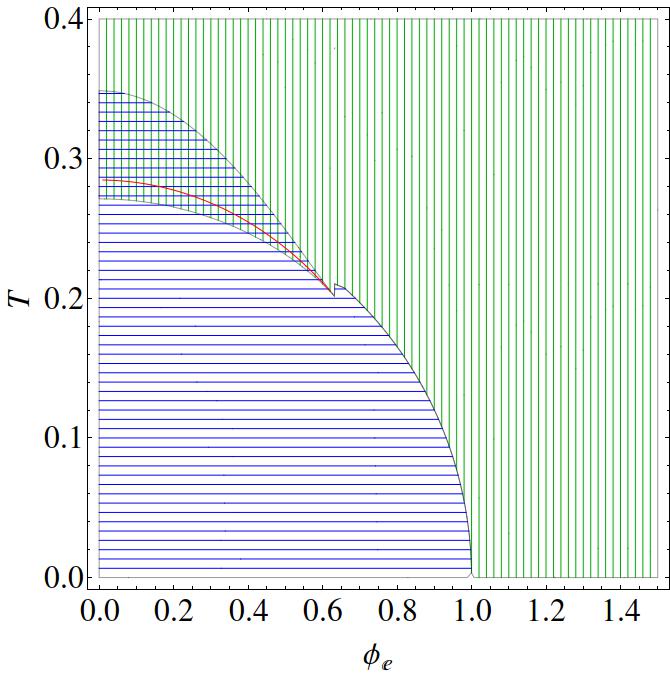}
		\caption{}
        \end{subfigure}
        \begin{subfigure}[b]{0.4\textwidth}
                \centering
                \includegraphics[width=\textwidth]{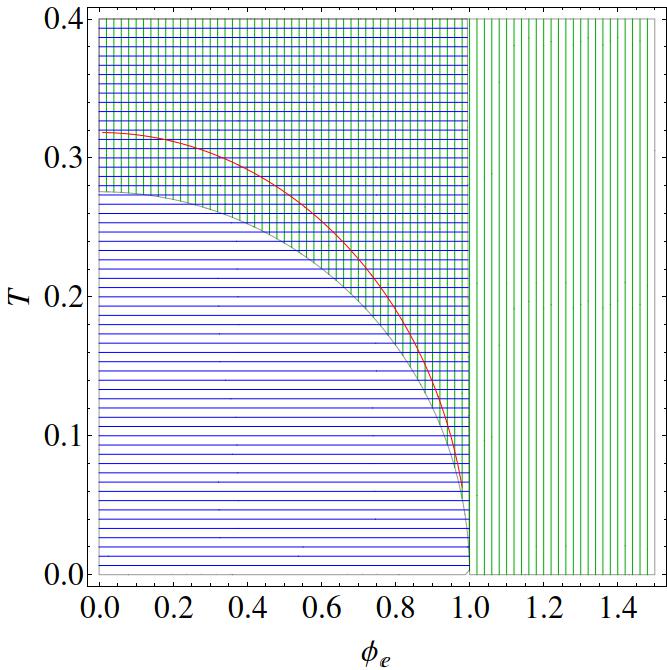}
                \caption{}
        \end{subfigure}
        \caption{Stability plots in $T-\Phi_E$ phase space for
          \textbf{(a)} $q_M<q_{M(c)}$ and \textbf{(b)}
          $q_M=0$. Horizontal blue lines correspond to Small BH
          solution stability region, whereas vertical green lines
          correspond to the stability region of Large BH. Red line
          corresponds to the phase transition.}\label{fig:stabphiT}
\end{figure}

\begin{figure}[h]
	\centering
        \begin{subfigure}[b]{0.4\textwidth}
                \centering
                \includegraphics[width=\textwidth]{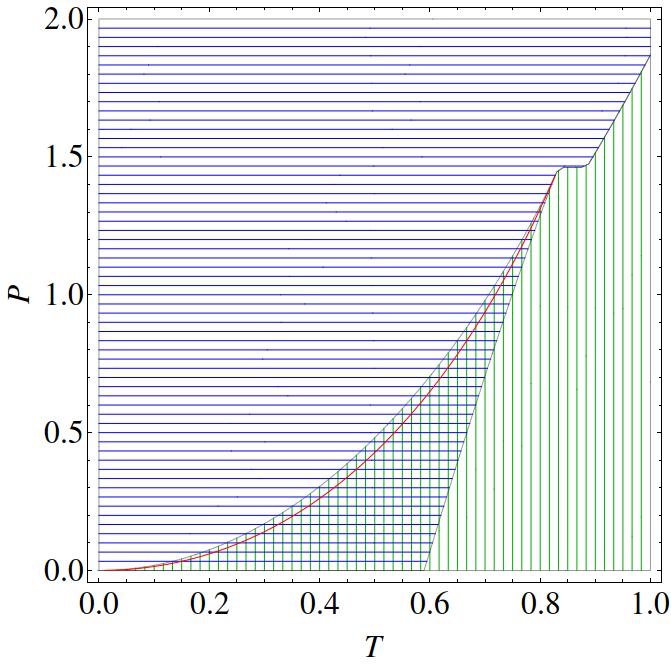}
		\caption{}
        \end{subfigure}
        \begin{subfigure}[b]{0.4\textwidth}
                \centering
                \includegraphics[width=\textwidth]{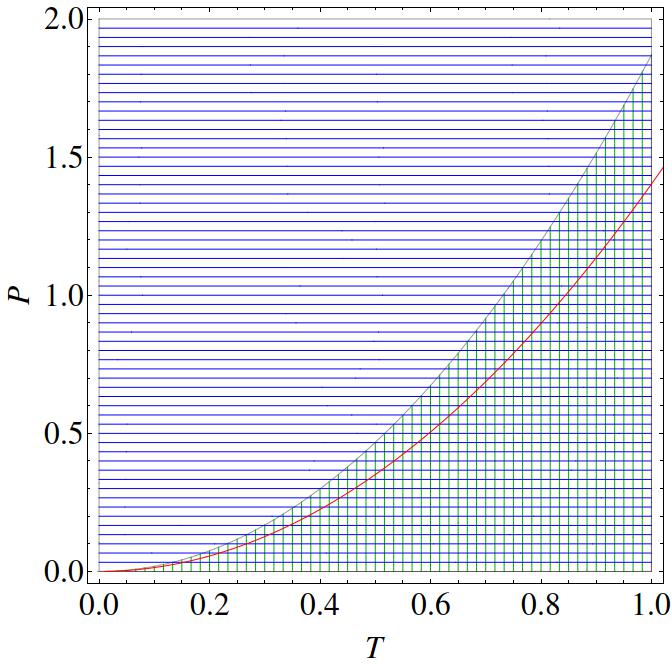}
                \caption{}
        \end{subfigure}
        \caption{Stability plots in $P-T$ phase space for \textbf{(a)}
          $q_M<q_{M(c)}$ and \textbf{(b)} $q_M=0$. Horizontal blue
          lines correspond to Small BH solution stability region,
          whereas vertical green lines correspond to the stability
          region of Large BH. Red line corresponds to the phase
          transition.}\label{fig:stabPT}
\end{figure}

We compute the zeroes of the Hessian of free energy $W=M-\Phi_E q_E
-TS$ with respect to $q_E$ and $r_+$ keeping $T$ and $\Phi_E$ fixed,
which gives us one condition (or bound) on the phase space. Positivity
of temperature will give another condition. These two conditions give
the following bound on the phase space.
\begin{equation}
  3q_M^2 + \frac{3 r_+^4}{b^2} - r_+^2(1-\Phi_E^2)>0, \quad\quad
  -q_M^2 + \frac{3 r_+^4}{b^2} + r_+^2(1-\Phi_E^2)>0 .
\end{equation} 
These are the same conditions we obtained by arguing positivity of
specific heat of the system (equation \ref{E:stabcp}). These
conditions can also be obtained from the relation
(\ref{E:physicalcond}): $P'(r_+)<0$, which is true for two solutions
- SBH and LBH. In figure (\ref{fig:stabphiT}) and (\ref{fig:stabPT})
We plot these stability lines and find that all values of $q_M$,
$\Phi_E$ and $T>0$ results in a stable solution.

\section{Concluding remarks} \label{conclu_remarks}

In the paper we have studied the phase diagram of dyonic black hole in
constant electric potential ensemble. An important thing to note here
is that the equation of state (\ref{E:EOS1}) does not contain the
global $AdS$ spacetime as a solution. The equation of state has two
stable solutions (SBH and LBH) and we find there is a first order
phase transition between them. However, when we take $q_M\ra 0$ limit,
the SBH evaporates to global $AdS$ and the SBH-LBH phase transition
boils down to usual Hawking-Page phase transition. Since we keep the
electric chemical potential constant the transition temperature
depends on the value of $\Phi_E$ \cite{Yamada:2006rx}. 

Our observation in section \ref{sec:magnetization} can be understood
from the electro-magnetic (EM) duality in the bulk. Since the bulk
spacetime is four dimensional, it admits an EM duality. The electric
charge can be interpreted as a magnetic charge in a dual frame and
vice versa. Therefore, our $\Phi_M$ in equation (\ref{thermoquantity})
can also be interpreted as $\Phi_E$ in the dual frame. Hence, one can
also have the same plots as shown in figure (\ref{fig:BM-TM}) and
(\ref{fig:BM-BM}) for $\Phi_E\ Vs. \ q_E$ or $\Phi_E\ Vs. \ T$ (see
figure (14) of \cite{pvcritic}).

 In \cite{pallav,Yamada:2006rx} the authors discussed different phases
 of boundary $CFT$ which are dual to different bulk solutions. They
 considered a unitary matrix model in the weak coupling side and
 showed that in the large $N$ limit, there exists three different
 saddle points which correspond to small, large and unstable black
 holes in the bulk. The key point of their observation is that in the
 canonical ensemble, the fixed electric charge constraint contributes
 an additional logarithmic term $\log (Tr U Tr U^{\dagger})$ involving
 the order parameter, to the gauge theory effective action. However,
 in our case, we have a constant magnetic field in the boundary
 theory. It is a good exercise to understand the effect of this
 constant magnetic field in the effective action of boundary theory.

\vspace{2cm}

%\newpage
\noindent
{\bf \large Acknowledgements}

We would like to thank Nabamita Banerjee, Umut Gursoy, Pratik Roy and
Dibakar Roychowdhury for many helpful discussions. SD would like to
acknowledge the hospitality of $Nikhef, \ Amsterdam$ where part of
this project was done. Finally we are also indebted to the people of
India for their unconditional support towards research and development
in basic science.

\bc
----------------------------------------
\ec

%\newpage

\vspace{2cm}

\appendix

\noindent
{\bf \Large Appendix}

\section{Background subtraction}
\label{backgroundsubtraction}
We calculate on-shell action by subtracting the contribution of an
extremal magnetically charged black hole with magnetic charge
$q_M$. The reason to choose this particular background depends on our
choice of ensemble (see \cite{myers}). The extremal black solution is
given by equation (\ref{Asol}) and (\ref{metric}) with $q_E\ra 0$
\begin{equation}\label{E:extremal1}
	f_e(r)=\left(
          1+\frac{r^2}{b^2}-\frac{2M_e}{r}+\frac{q_M^2}{r^2} \right) .
\end{equation} 

We consider that the extremal black hole has a constant electric
potential which is same as $\Phi_E$ of the black hole. The black hole
has a horizon radius $r_e$, given by
\begin{equation}
	f_e(r_e)=\left( 
1+\frac{r_e^2}{b^2}-\frac{2M_e}{r_e}+\frac{q_M^2}{r_e^2} \right)=0 .
\end{equation} 
The extremality condition is given by
\begin{equation}\label{E:extremalcond}
	r_e^2+\frac{3r_e^4}{b^2}=q_M^2 .
\end{equation} 
Therefore, the on-shell action of background turns out to be
\begin{equation}
  I_E=\frac{1}{16 \pi}\int^{R}_{r_e}
  d^4x\sqrt{g}\left(F^2+\frac{6}{b^2}\right)=\frac{\beta'}{4} 
\left[\frac{-2q_M^2}{R}
    +\frac{6R^3}{3b^2}-\frac{-2q_M 
      2}{r_e}-\frac{6r_e^3}{3b^2}
  \right] 
\end{equation}
where, $\beta'$ is the radius of the Euclidean time circle of
background. $\beta'$ can be obtained by identifying the asymptotic
boundary geometry of the black hole spacetime and extremal black hole
spacetime
$$\displaystyle \beta\sqrt{g_{\tau\tau}}=\beta'\sqrt{g^{e}_{\tau\tau}}
 .$$
Thus, we get
\begin{equation}
	\beta ' = \beta \left(1-\frac{(M-M_e)b^2}{R^3}\right) .
\end{equation}
Subtracting the contribution of extremal black hole on-shell action
from that of black hole we finally get,
\begin{equation}
  I_{onshell}=I_{BH}-I_E=\frac{\beta}{4}\left[ -\Phi_E^2r_+ +
    \frac{3q_M^2}{r_+} -\frac{r_+^3}{b^2} +r_+ -4r_e
    -\frac{8r_e^3}{b^2} \right] .
\end{equation}
Hence, the free energy is given by,
\begin{equation}\label{E:firstfreeenergyapp}
	W=\frac{I}{\beta}=\frac{1}{4}\left[ -\Phi_E^2r_+
          +\frac{3q_M^2}{r_+} -\frac{8\pi Pr_+^3}{3} +r_+ -4r_e -
          \frac{64\pi Pr_e^3}{3} \right] . 
\end{equation}
This renormalization prescription tells us that all the thermodynamic
quantities we measure (for example energy, volume, charges etc.) are with
respect to an extremal magnetic black hole background.

Once we have free energy $W(T,\Phi_E,q_M,P)$ at hand, all the
thermodynamical variables can be computed easily. We quote the results
as follows:
\begin{align}\label{thermoquantity2}
  q_E 			&=-\frac{\partial W}{\partial \Phi_E}=\Phi_Er_+ ;
\nonumber\\
  \Phi_M-\Phi_M^{(e)}	&=\frac{\partial W}{\partial 
q_M}=\frac{q_M}{r_+}-\frac{q_M}{r_e} ; \nonumber\\
  S &=-\frac{\partial W}{\partial T}=\beta^2\frac{\partial W}{\partial
    \beta}=\frac{1}{4}4\pi
  r_+^2=\frac{A_H}{4} ; \nonumber\\
  V-V_e.  &=\frac{\partial W}{\partial
    P}=\frac{4\pi}{3}r_+^2-\frac{4\pi}{3}r_e^2 .
\end{align} 
where $\Phi_M$ is the chemical potential corresponding to magnetic 
charge $q_M$. Note that, volume and magnetic potential is measured 
with respect to the extremal BH background. Using equation 
(\ref{E:freeenergy0}) we can calculate 
the enthalpy $E$ of the BH given by:
\begin{equation}
	E=W+TS+\Phi_E q_E =M-M_e .
\end{equation} 

\section{Critical points and $\M$ vs. $B$ plots}\label{App:MT-BM}

$W$ in equation (\ref{E:freeenergyBM}) can be plotted against $T$ for
varying $B$. Since $B$ is identified with $q_M$, the plot is the same
as figure (\ref{fig:q0}). This time Branch-1 corresponds to high
magnetization (Small black hole), and Branch-3 corresponds to low
magnetization (Large black hole). We note the key points of
Small-Large BH phase transition in this language:

Critical Curve is given by:
\begin{equation}
	T_c= \frac{2}{\pi \sqrt{6}b}(1-\Phi_{E(c)}^2)^{1/2}
\end{equation} 
or correspondingly:
\begin{equation}
	B_c =\pm\frac{1}{6b}(1-\Phi_{E(c)}^2)
\end{equation} 
while, Magnetization $\mathcal{M}$ at critical point(s) is given by:
\begin{equation}
  \mathcal{M}_c =\mp\frac{b^2}{\sqrt{6}}(1-\Phi_{E(c)}^2)^{1/2}.
\end{equation} 
\begin{figure}[h]
        \begin{subfigure}[b]{0.5\textwidth}
                \centering
                \includegraphics[width=\textwidth]{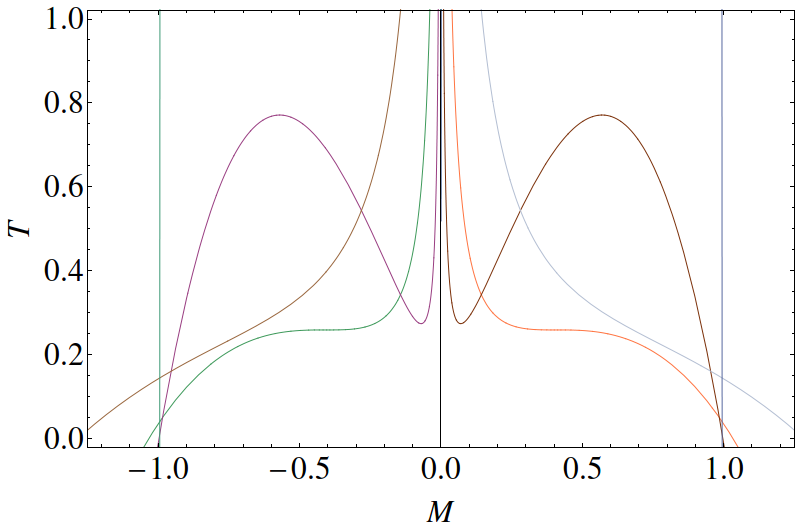}
		\caption{}\label{fig:BM-TM}
        \end{subfigure}
	\begin{subfigure}[b]{0.5\textwidth}
                \centering
                \includegraphics[width=\textwidth]{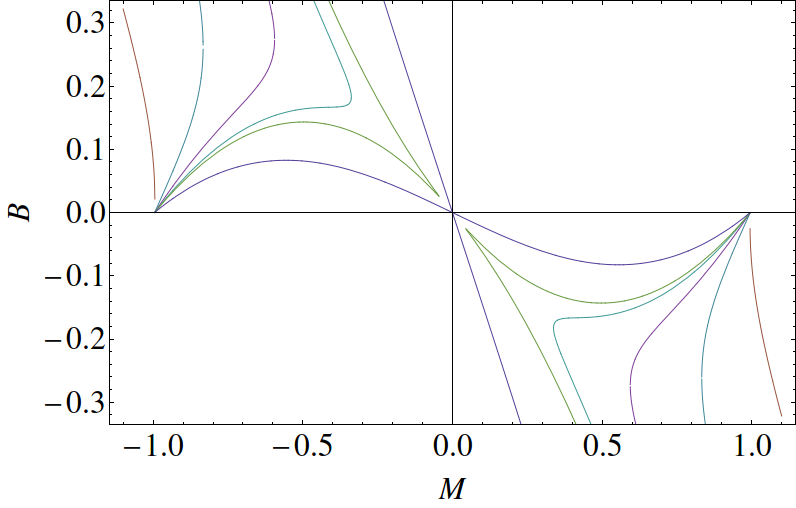}
		\caption{}\label{fig:BM-BM}
        \end{subfigure}
	\caption{\textbf{(a)} $T-\mathcal{M}$ graph for varying $B$
          ($\Phi_E=0.1$). In order of the graphs near the $T$ axis for
          small values of $\mathcal{M}$, the leftmost is for the
          highest positive $B$, and the rightmost for highest negative
         (magnitude) $B$. Criticality occurs in Green ($\M <0$) and Light
          Orange ($\M >0$) graphs ($B=\pm 0.165$). \textbf{(b)}
          $B-\mathcal{M}$ graph for varying $T$ ($\Phi_E=0.1$). In
          order of the graphs in the second quadrant, the leftmost is
          for the lowest $T$ and the rightmost for highest $T$. Curves
          in the fourth quadrant are the segments of the same
          graphs. Criticality occurs at $T=0.258$.}
\end{figure}
A thing to note here will be the Nucleation temperature, $T_N$. As
$B\rightarrow 0$, $T_N$ saturates to some maximum value
$T_{N(M)}$. Same is evident in figure (\ref{fig:BM-BM}), (\ref{fig:BM-TM})
 ($T_{N(M)} \approx 0.3$). It is the temperature when a curve (In
 figure \ref{fig:BM-TM}) touches the origin. At $T>T_{N(M)}$
 (representative blue graph), $B=0$ has 3 solutions.
 $T_{N(M)}$ will be given by the minima of $T-\mathcal{M}$ graph as
 $B\rightarrow 0$, which using equation \ref{E:tempBM} is:
\begin{align}\label{E:maxnuc}
	T_{N(M)}	&=\frac{\sqrt{3}}{2\pi b}\sqrt{1-\Phi_E^2} .
\end{align} 
Similar to the nucleation temperature, coexistence temperature $T_o$
also reaches its maximum value $T_{o(M)}$ as $B\rightarrow 0$. It can
be easily found by finding root of $W$ which gives:
\begin{equation}\label{E:maxcoex}
	T_{o(M)}=\frac{1}{\pi b}\sqrt{1-\Phi_E^2} .
\end{equation} 
The coexistence point for the phase transition can also be translated
similarly demanding $W$ to be same for two phases, which will give
corrected $B-\mathcal{M}$ and $T-\mathcal{M}$ plots. Coexistence curve (and 
plane) is however found to be like figure (\ref{fig:coexPBM}).
\begin{figure}[h]
	\centering
        \includegraphics[width=.5\textwidth]{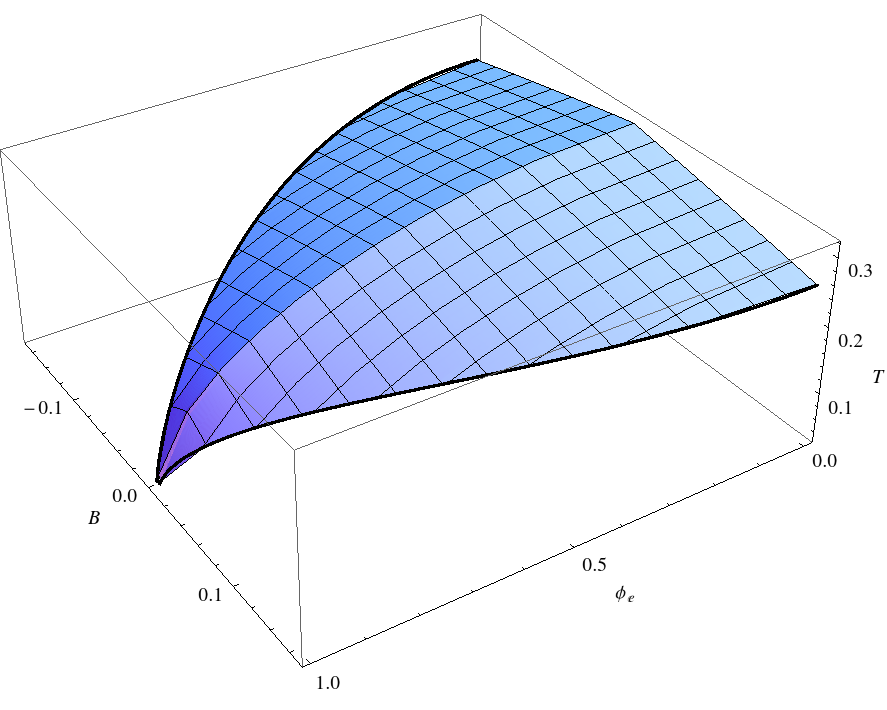}
        \caption{\textbf{(b)} Black Curve is the Critical Curve in
          $T-B-\Phi_E$ plane. The manifold is the coexistence
          plane. Above this plane lies the Small BH, and below it Large 
BH.}\label{fig:coexPBM}
\end{figure}

\section{Magnetization from the minima of free
  energy}\label{App:Ferro} 

As we have seen that our system possesses a `ferromagnetic' type
behaviour in $B\ra0$ limit. In this appendix we investigate possible
minima of the free energy as function of temperature. We write the
free energy from equation (\ref{E:freeenergy0}):
\begin{align}
  W	&=M-TS-\Phi_E q_E, \\
  &=-\frac{1}{2}\lb\frac{B}{\mathcal{M}}\rb(1-\Phi_E^2)b^4-\frac{B\M}{2}-\frac{
    b^{10}}{2}\lb\frac{B}{\mathcal{M}}\rb^3-\pi
  Tb^8\lb\frac{B}{\mathcal{M}}\rb^2. \label{E:freenergylandauBM}
\end{align}
Using positive definiteness of $r_+$ we can write
\begin{align}
  W
  &=\frac{1}{2}\left\vert\frac{B}{\mathcal{M}}\right\vert(1-\Phi_E^2)b^4+\frac{|B\M|}{2}+\frac{
    b^{10}}{2}\left\vert\frac{B}{\mathcal{M}}\right\vert^3-\pi
  Tb^8\left\vert\frac{B}{\mathcal{M}}\right\vert^2. \label{E:freeenergybyf}
\end{align}
$\partial W/\partial \mathcal{M}=0$ will give us the back the equation
of state (equation \ref{E:tempBM}), and we will recover figure
(\ref{fig:BM-TM}) and (\ref{fig:BM-BM}). To find the stable solutions
of $\mathcal{M}$ in $B\ra0$ limit, we study the extremas of $W$ using
(\ref{E:freenergylandauBM}). For $\displaystyle
r_+=\left.-b^4\frac{B}{\M}\right\vert_{B\ra0}=0$ we will have:
\begin{equation}
	\mathcal{M}=\pm b^2\sqrt{(1-\Phi_E^2)} .
\end{equation} 
Whereas for $r_+\neq 0$ the only solution is:
\begin{equation}
	\mathcal{M}=0 \quad \text{if} \quad T\geq T_{N(M)}= 
\frac{\sqrt{3}}{2\pi b}\sqrt{\left(1-\Phi_E ^2\right)} .
\end{equation} 
First one is the background $AdS$ solution which has some net finite
magnetization even without any magnetic field.  The second solution
however corresponds to large black hole. The bound on the temperature
is nothing but the maximum nucleation temperature (equation
\ref{E:maxnuc}), above which (for $B=0$) there will always exist large
black hole solution with radius
\begin{equation}\label{E:r+ofLBH}
	r_{+(LBH)}=\frac{2\pi b^2}{3}\lb T+\sqrt{T^2-T_{N(M)}^2}\rb .
\end{equation} 

To argue the dominance of the above solutions, we express $W$ in the
convenient form of $r_+$ in $B\rightarrow 0$.
\begin{align}
	W	&=\frac{r_+}{2b^4}\left[(1-\Phi_E^2)b^4+\mathcal{M}^2+b^2r_+^2
	-2\pi Tb^4r_+\right]\label{E:freeenergybyB3} .
\end{align}
For $AdS$ solution, $\mathcal{M}=\pm\sqrt{1-\Phi_E^2}$ and $r_+=0$, so
the free energy vanishes. For the large black hole solution the 
free energy is given by
\begin{align}\label{E:freeenergyLBH}
  W_0
  &=\frac{1}{2}r_{+(LBH)}\left[(1-\Phi_E^2)+\frac{r_{+(LBH)}^2}{b^2}-2\pi
    Tr_{+(LBH)}\right].
\end{align} 
When large black hole solution dominates over the global $AdS$ then
$W<0$. Using equation (\ref{E:r+ofLBH}) which boils down to:
\begin{equation}
	T>T_{o(M)}=\frac{1}{\pi b}\sqrt{1-\Phi_E ^2} .
\end{equation} 
This bound is the maximum coexistence temperature (equation
\ref{E:maxcoex}), the temperature at which $AdS$ and large black hole
coexist at $B=0$. Above $T_{o(M)}$, $\mathcal{M}=0$ solution
dominates, while below it $\mathcal{M}=\pm\sqrt{1-\Phi_E^2}$
dominates. But the latter two have the same free energy (equation
\ref{E:freeenergyLBH}). Thus system has to spontaneously choose
between these two solutions. As soon as $B$ is applied one of the
solutions disappears\footnote{Since $B$ and $\M$ must have opposite
  signs, one of the solutions of $\M$ vanishes depending on the sign
  of $B$.}, and system has a preferred direction. This behaviour has
similarity with ferromagnetic system.

\begin{figure}[h]
        \begin{subfigure}[b]{0.5\textwidth}
                \centering
                \includegraphics[width=\textwidth]{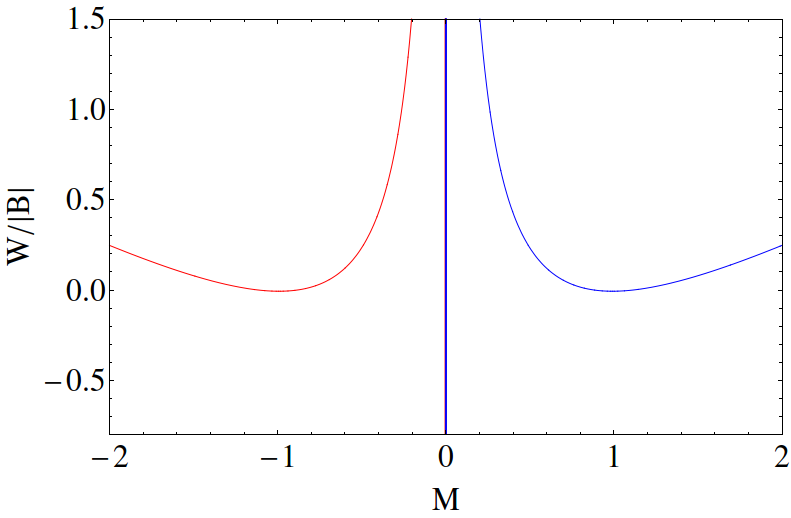}
                \includegraphics[width=\textwidth]{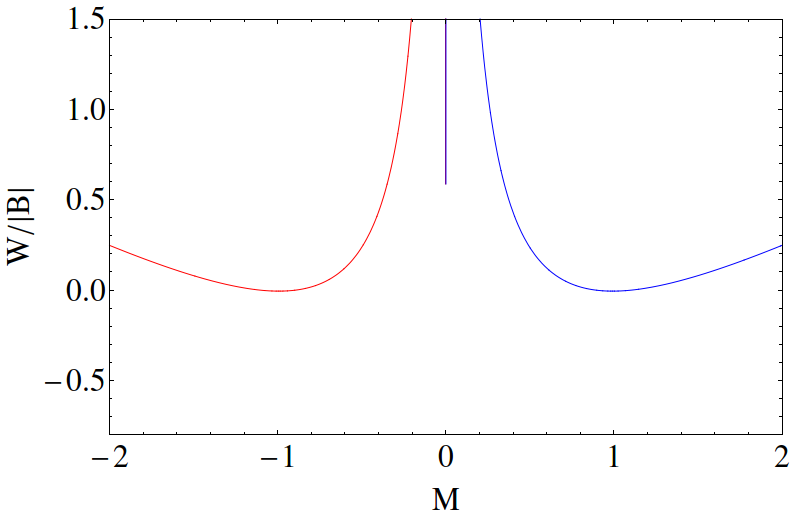}
		\caption{}
        \end{subfigure}
        \begin{subfigure}[b]{0.5\textwidth}
                \centering
                \includegraphics[width=\textwidth]{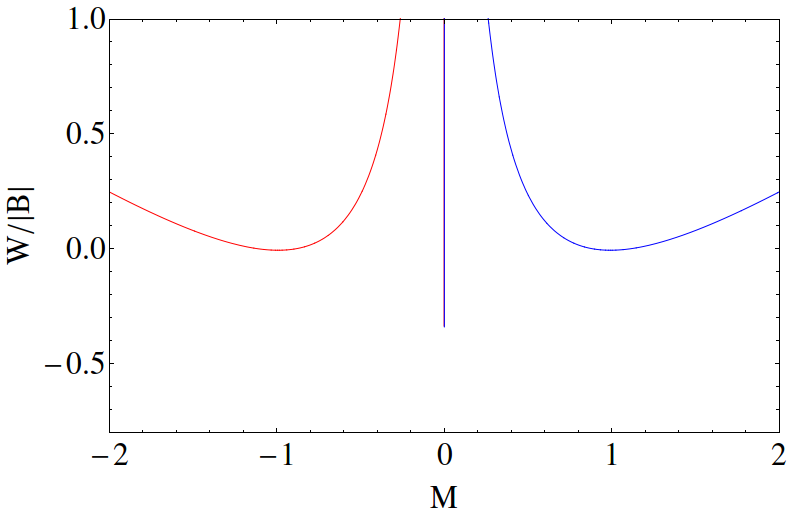}
                \includegraphics[width=\textwidth]{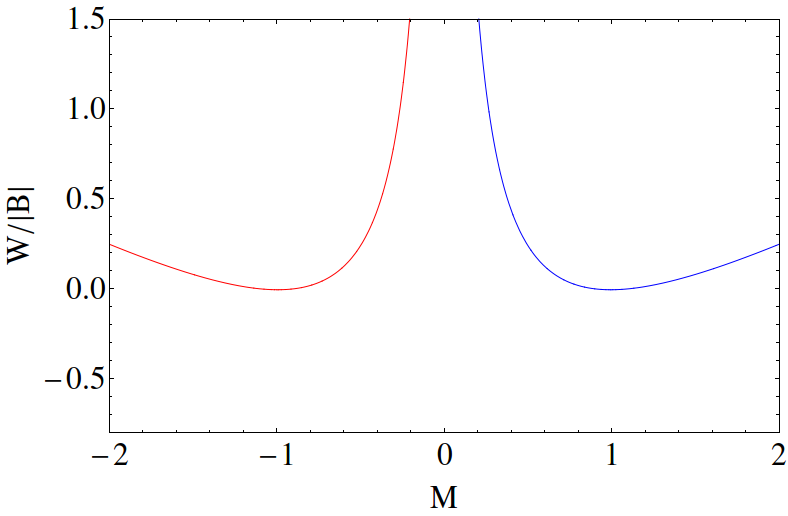}
                \caption{}
        \end{subfigure}
        \caption{$W/|B|-\M$ graph for varying $T$ ($\Phi_E=0.1$) as
          $B\rightarrow 0$. Red curve is for $B\rightarrow 0^+$ while
          Blue is for $B\rightarrow 0^-$. In the first and second
          graph, $T>T_{o(M)}$ (coexistence temperature) and we see
          that the thermodynamics is governed by $\mathcal{M}= 0$
          (Large BH) solution. As $T$ is dropped below $T_o$ (third
          graph), system spontaneously chooses between the two
          available solutions of $\mathcal{M}=\pm \sqrt{1-\Phi_E^2}$
          (Small BH or background $AdS$) as both are equally
          likely. Further when $T$ is decreased even below $T_{N(M)}$,
          the $\mathcal{M}=0$ solution vanishes altogether (fourth
          graph).}\label{fig:BM-MB}
\end{figure}

Notice that the free energy is always zero in $B\ra 0$ limit (except
for $\M=0$). However we can have some insight if we plot $W/|B|$
instead, which keeps the extremas of $W$ unaltered\footnote{We cannot
  divide by $B$ as its sign has a dependence on $\M$ due to positivity
  of $r_+$} (figure \ref{fig:BM-MB}).

Ideally one would be tempted to Taylor expand equation
(\ref{E:freeenergybyf}) near $\M=0$, and construct a Landau-type model
to see the symmetry breaking explicitly. However this is not feasible
in our case, as $W/|B|$ is discontinuous and diverging at $\M=0$ due
to the inclusion of $|\M|$, and thus the Taylor expansion breaks down.

\iffalse For $B\ra0$ however, we know that $r_+$ is either 0 (for SBH)
or a constant given by equation (\ref{E:r+ofLBH}) (for LBH). Thus we
can write it in terms of $\Delta T = T-T_{o(M)}$ as:
\begin{align}
  r_+= r_{+(LBH)} \ \tilde\theta(\Delta T,\M) \quad\quad s.t. \quad
  \lim_{B\ra0}\tilde\theta(x,\M)\ra \theta(x) =
  \left\{ \begin{array}{l l}
      1 & \quad x>0,\\
      0 & \quad x<0.
				\end{array} \right.
\end{align}
Also using equation (\ref{Mvalue}),
\be
	\frac{\partial \tilde\theta}{\partial \M}=-\frac{\tilde\theta}{\M}.
\ee
Therefore in $B\ra0$ limit we our free energy reduces to:
\begin{align}
  \frac{W}{|B|}=\frac{|\M|}{2}+\frac{1}{2|\mathcal{M}|}\left[b^4(1-\Phi_E^2)+\theta(\Delta
    T)K\right], \label{E:freeenergybyfr}
\end{align}
\be
	K(T,\Phi_E)=b^{2}r_{+(LBH)}^2-2\pi Tb^4r_{+(LBH)}
\ee
%In $B\ra0$ limit we exchange $B$ for $r_+$, bearing in mind that it is still a function of $\M$. 
%Therefore we can write it as:

%Thus $f$ will reduce to:
%and using $\displaystyle \frac{\partial r_+}{\partial \M}=-\frac{r_+}{\M}$:
%\begin{align}
%	\frac{\partial f}{\partial \M}\ra\frac{|\M|}{2\M^3}\left[\lb \M^2 - b^4(1-\Phi_E^2)\rb-\theta(\Delta T)\lb 3b^2r_{+(LBH)}^2 - 4 \pi Tb^4 r_{+(LBH)} \rb\right]
%\end{align}
\fi

\section{Details of diamagnetic and paramagnetic
  phases}\label{A:detdiapara}

Equation (\ref{E:chiB}) gives behaviour of $\chi$ with $B$. Thus if we
study $\chi$ at a constant magnetic field, system will be diamagnetic
if:
\begin{equation}
	3b^{10} B^2+3\mathcal{M}^4 - b^4\mathcal{M}^2\left(1-\Phi_E^2\right)<0
\end{equation} 
or 
\begin{equation}
	3b^{10} B^2+\mathcal{M}^4 - b^4\mathcal{M}^2\left(1-\Phi_E^2\right)>0
\end{equation} 
and otherwise paramagnetic. That is diamagnetic solution is given by:
\begin{equation}
	\mathcal{M}< \frac{b^4}{2} 
\left(\left(1-\Phi_E^2\right)-\sqrt{\left(1-\Phi_E^2\right)^2-12 b^2 B^2}\right)
\end{equation} 
\begin{equation}
	\frac{b^4}{6} 
\left(\left(1-\Phi_E^2\right)-\sqrt{\left(1-\Phi_E^2\right)^2-36 b^2 
B^2}\right)< \mathcal{M} < \frac{b^4}{6} 
\left(\left(1-\Phi_E^2\right)+\sqrt{\left(1-\Phi_E^2\right)^2-36 b^2 B^2}\right)
\end{equation} 
\begin{equation}
	\mathcal{M}> \frac{b^4}{2} 
\left(\left(1-\Phi_E^2\right)+\sqrt{\left(1-\Phi_E^2\right)^2-12 b^2
    B^2}\right) .
\end{equation} 
The middle one is however the unstable phase. If,
\be
	|B|>B^*=\frac{1}{2\sqrt{3} b}\left(1-\Phi_E^2\right)
\ee
the system will always be diamagnetic regardless of temperature (figure 
\ref{fig:chiTB1}). However if,
\be
	B^*>|B|>|B_c|
\ee
the system will have three phases: diamagnetic, paramagnetic and another 
diamagnetic (figure \ref{fig:chiTB2}).

If the magnetic field is even below $B_c$, system shows the full
feature of 5 phases. But as we go below the critical point, we know that there comes an 
unstable branch which is omitted by the Maxwell's equal area law (figure 
\ref{fig:cortm}). While we do so, a section of the small and large BH is also 
omitted. As a result below a particular magnetic field $B^\#$, the paramagnetic 
phase of large BH gets cutout and the only stable solution of large BH is 
diamagnetic phase. $B^\#$ will be given by the point where the coexistence 
magnetic field is same as the $\chi=0$ point for a $B-\mathcal{M}$ isotherm. 
For $\Phi_E=0$ and $b=1$ we find $B^\#\approx 0.154$ whereas $B_c=0.167$.

\end{document}